\documentclass[draft]{agujournal2019}
\usepackage{url} 
\usepackage{lineno}
\usepackage[inline]{trackchanges} 
\usepackage{amsmath, amssymb}
\usepackage{soul}

\newcommand{\bvec}[1]{\mathbf{#1}}
\newcommand{\pa}{\partial}
\newcommand{\Vlasiator}{\texttt{Vlasiator}}

\draftfalse

\journalname{JGR: Space Physics}

\begin{document}

%
%


\title{Geoelectric Field Caused by Flux Transfer Events in an Ionosphere-Coupled Vlasiator Simulation}

%
%




\authors{K.~Horaites\affil{1,2}, M.~Alho\affil{3}, Y.~Pfau-Kempf\affil{4}, U.~Ganse\affil{3}, A.~Workayehu\affil{3}, J.~Suni\affil{3}, F.~Tesema\affil{5}, L.~Juusola\affil{3,6}, G.~Cozzani\affil{7}, S.~Hoilijoki\affil{3},  I.~Zaitsev\affil{3},  S.~Kavosi\affil{8}, M.~Palmroth\affil{3,6}}

\affiliation{1}{CIRES, University of Colorado at Boulder, Boulder, CO, USA}
\affiliation{2}{Space Science Institute, Boulder, CO, USA}
\affiliation{3}{Department of Physics, University of Helsinki, Helsinki, Finland}
\affiliation{4}{CSC – IT Center for Science, Espoo, Finland}
\affiliation{5}{Department of Physics and Technology, University of Bergen, Bergen, Norway}
\affiliation{6}{Finnish Meteorological Institute, Helsinki, Finland}
\affiliation{7}{LPC2E, CNRS/CNES/University of Orléans, Orléans, France}
\affiliation{8}{Air Force Research Lab, Kirtland AFB, NM, USA}




\correspondingauthor{Konstantinos Horaites}{konstantinos.horaites@colorado.edu}




\begin{keypoints}
\item A Flux Transfer Event (FTE) can be segmented into several flux ropes, which are separated along the FTE axis by 3D magnetic null points. 
\item FTEs launch pulsed field-aligned currents towards Earth,  which propagate along the magnetic field lines at the Alfv\'en speed.  
\item Along with the pulsed currents, we observe 
rotational geoelectric fields that move from the noon meridian around the auroral oval.
\end{keypoints}

%
%

%
%


\begin{abstract}
We report on the relationship between flux transfer events (FTEs) at Earth's magnetopause and the geoelectric field that is induced near the FTEs' magnetic footpoints. We study this system using the global hybrid-Vlasov code \Vlasiator, which has recently been extended to model ionospheric physics. 
We also highlight the significance of 3D magnetic null points, which in our simulation can separate the FTEs into multiple flux ropes. 
Near the null points, the coiled FTE magnetic field lines are rerouted towards Earth, so that the magnetic footpoints are planted near the Region~1 ionospheric current system. The helicities of the flux ropes are organized by the $y$-component (GSE) of the magnetic field at the Earth's magnetopause. This occurs in our simulation due to the absence of a $y$-component of the interplanetary magnetic field, which would normally supply the FTE guide field that determines the helicity. We observe Alfv\'enic, Earthward-flowing field-aligned currents generated near the magnetopause that correlate with the passage of FTEs nearby. These pulses of current coincide with the formation of rotational geoelectric field structures, that appear near the noon meridian and propagate around the auroral oval towards the nightside.
\end{abstract}

\section*{Plain Language Summary}
The geoelectric field is a phenomenon that generates electrical currents in Earth's surface, whose origins can be traced to outer space.
The geoelectric field is important because it can drive unwanted ``geomagnetically induced currents'' (GICs)  in long conductors on the ground, which may cause blackouts in the power grid or disturb long-range telecommunications.
The geoelectric field and the attendant GICs are formed indirectly when electrical currents in space flow down to the ionosphere, a high-altitude layer of Earth's atmosphere.
The ionospheric currents then spread out horizontally in order to close the circuit, which then causes secondary currents to sympathetically flow in the ground. 
 In this study, we use computer simulations of the Earth's space environment to pinpoint the origins of the geoelectric field.
 Specifically, we show how ``flux transfer events'' (FTEs)---dense gaseous structures that form during explosive reconfigurations of the magnetic field---can cause geoelectric fields on Earth. 
 The FTEs appear at the boundary between Earth's magnetic field and interplanetary space, but are nonetheless connected to Earth's surface. 
 By using simulations that resolve both Earth's ionosphere and the outer space environment, we can track the causal chain that links the distant FTEs to the geoelectric field on Earth.

%
%

%


%
%
%
%

\section{Introduction}\label{intro_sec}


The geoelectric field is a major concern in the study of space weather. The geoelectric field is induced by magnetic field fluctuations at Earth's surface, driving Ohmic ``telluric currents'' in the ground itself and ``geomagnetically induced currents'' (GICs) in long grounded conductors. The GICs can disrupt and even permanently damage electrical power grids \cite<e.g.>{abda20}.
Observational studies of these induced currents have primarily considered the interaction between the Earth's conducting crust and surface magnetic fields \cite<e.g.,>{pulkkinen06}. This is because the magnetic field time derivative is easier to directly measure than the geoelectric field it induces.
A thorough description of the ground signatures must also account for how they are ultimately caused by physical processes in outer space.
In this study, we consider how Flux Transfer Events (FTEs) associated with reconnection in Earth's dayside magnetosphere can lead to a geoelectric field at the surface. We analyze the behavior of a global hybrid-kinetic simulation of Earth's magnetosphere, performed with the \Vlasiator\ code \cite{palmroth25_vlasov_methods_living_review, ganse23}. \Vlasiator\ has recently been extended to model the Earth's ionosphere \cite{ganse_2025_vlasiator_ionosphere}, enabling a detailed investigation of the link between FTEs and geoelectric fields and its space weather ramifications. 


The surface magnetic field fluctuations responsible for the geoelectric field derive from the time-varying currents flowing throughout the ionosphere-magnetosphere system. For an observer on the ground, the primary contribution to the local magnetic field comes from the ionospheric currents, which flow overhead at an altitude of $\sim$100~km. 
Incoming field-aligned currents (FACs), originating from the magnetosphere, are closed horizontally through this conductive ionospheric layer before exiting vertically. The ionospheric currents thus provide a way to close the global system without local accumulation of charge. The divergence-free  component of the ionospheric current is most responsible for the ground magnetic field perturbations, by Fukushima's theorem \cite{fukushima76}. By the same token, an enhancement of the inflowing FACs is likely to be associated with an intensification of the ionospheric currents and the resulting geoelectric fields. Understanding the origins of the FACs, which are generated in outer space, is therefore essential to explaining how currents are induced on the ground. 


A Flux Transfer Event is a magnetic reconnection phenomenon that occurs at the dayside magnetopause and is believed to play a role in transmitting FACs to Earth.  The first observations of FTEs \cite{haerendel78, russellelphic78} revealed that during periods of southward-oriented interplanetary magnetic field (IMF), the reconnection near the magnetic equator is transient rather than persistent.
These bursts of reconnection change the magnetic field topology, producing long flux ropes that extend across the magnetopause \cite<e.g., >[]{eggington22}.
The motion of these flux ropes is influenced by global flow of the solar wind around the magnetopause, as well as the reconnection outflow jets. 
The bulk plasma flow tends to carry the flux ropes towards the Earth's poles, where they may reconnect again upon encountering the magnetic cusps \cite{omidi_2007_cusp_reconnection}.
As considered in \citeA{southwood85, southwood87}, FACs produced by FTEs should be carried by shear Alfv\'en waves down to Earth, developing distinctive Pedersen and Hall current patterns upon reaching the ionosphere. 
The ionospheric footprints of the FTEs are predicted to be magnetic downmappings of ``holes'' in the magnetopause surface \cite{crooker_1990_mapping_FTEs_to_ionosphere}.
As noted in \citeA{glassmeierstellmacher96}, the Alfv\'enic wave pulses and ionospheric signatures are difficult to identify due to the scarcity of simultaneous ground- and space-based observations along the same field line. 
Although some progress has been made in establishing the ground signatures of FTEs \cite<e.g.,>[]{elphic_1990_FTE_magnetopoause_ionosphere, daum_2008_FTE_m-i_coupling_observations_and_MHD_simulations}, the lack of coordinated ground-space observations has remained an issue.

The semantics of the phrase ``Flux Transfer Event'' is highly dependent on the context and a given author's preferences. Among its meanings, an ``FTE'' may refer to an isolated temporal process of magnetopause reconnection, the quantified erosion of the magnetospheric flux resulting from a reconnection event, 
or a physical plasmoid or flux rope that is formed by the reconnection \cite<see>[and discussion therein]{russellelphic78, fear_2017_flux_transferred_by_FTEs, pfau-kempf_2025_fte_evolution_vlasiator}. 
At our discretion, in this work we will consider FTE as a physical object, i.e. a coherent structure of bundled-up magnetic field lines that results from the magnetopause reconnection. We refrain from equating FTEs with flux ropes because, as we show in our analysis, an FTE can be comprised of one or more flux ropes that are adjoined along a seemingly-continuous axis.


Several numerical studies have addressed the ionospheric impact of FTEs. \citeA{omidi_2007_cusp_reconnection} performed 2.5-dimensional hybrid particle-in-cell (PIC) simulations, and found that the passage of FTEs is linked to transient increases in the Earthward flux of energetic ions. They also found that the ionospheric signatures  of the precipitating ions are reminiscent of so-called ``poleward-moving auroral forms'' (PMAFs).  \citeA{grandin23} corroborated this picture using 3-dimensional hybrid-Vlasov \Vlasiator\ simulations, highlighting the importance of the cusp-FTE and matching the predicted ionospheric fluxes with in-situ observations, made in low-Earth orbit by the Defense Meteorological Satellite Program spacecraft. In a similar vein, \citeA{daum_2008_FTE_m-i_coupling_observations_and_MHD_simulations} studied a conjunction between the ground-based SuperDARN auroral radar and the magnetospheric in-situ Cluster observations. The authors modeled this event with the BATS-R-US 3-dimensional MHD code, using the observed solar wind driving conditions at L1 to drive the simulation, and their simulations predicted a combined azimuthal-poleward motion of the FTE footpoints.  The FTE motion was studied on a global scale in \cite{pfau-kempf_2025_fte_evolution_vlasiator}, using the same 3-dimensional \Vlasiator\ simulation investigated here. The authors traced the FTE volumes, and found that some FTEs---whose axes are aligned with the plasma flow and whose magnetic field does not shear significantly with the magnetic field of the magnetospheric lobes---can propagate past the Earth, continuing for tens of Earth radii before eventually reconnecting in the magnetopause flanks. The authors found that FTEs that are located farther behind the Earth (more anti-sunward) tend to have footpoints farther from the noon meridian.


One recent study, \citeA{paul23}, has supplied a broad overview of other FTE impacts in the context of a global magnetospheric code named \texttt{MagPIE} (``Magnetosphere of Planets with Ionospheric Electrostatics''). The simulations evolve the magnetosphere according to a set of resistive-MHD equations, which are then coupled with a height-integrated model ionosphere. In a detailed case study, the authors identified FTEs in each hemisphere of the simulation, and plotted several important quantities around the magnetic field's ionospheric footpoints. The authors reported that FACs are generated when an FTE reconnects with the Earth's magnetic cusp. The FACs (as well as the ionospheric potential and vortex flow) at the FTE footpoints exhibit a ``tripolar'' structure, in contrast with the simple bipolar scenario outlined in \citeA{southwood87}. \citeA{paul23} also reported that the polar projection of the FACs may also appear as a pair of I- and U-shaped regions of oppositely directed currents. These structures are concentrated around the ionospheric open-closed boundary, and they move azimuthally around the auroral oval towards the nightside. This motion was reportedly due to the changing site of the FTE-cusp interaction. Rotational ionospheric velocity field patterns, so-called ``traveling convection vortices'' (TCVs), were observed to be generated in the vicinity of the azimuthally-moving FAC signatures.


In the present work, we further the understanding of the space weather impacts of FTEs by considering the electromagnetic signatures they produce on the ground.
We use an ionosphere-coupled \Vlasiator\  simulation of the global magnetosphere (section~\ref{simulation_sec}) to study this process. 
Our approach is similar to \citeA{paul23}, but in our analysis we concentrate on the strength and structure of the geoelectric field, which is closely related to the telluric currents and GICs. 
The geoelectric field is induced by fluctuating magnetic fields, mediated by the influence of the conducting crust.
Although we adopt a simplified model of the ground conductivity, this analysis still sets an important baseline expectation for the geoelectric signatures resulting from the FTEs. 
We also investigate the topology of the FTE flux ropes, including their helicity, and discuss how these characteristics correspond with ionospheric signatures. We additionally demonstrate the correlation between the ground signatures and the timing of the FTEs' passage across the magnetopause, characterizing the Alfv\'enic field-aligned current pulse that propagates towards Earth.

The remainder of the paper is organized as follows. Section~\ref{simulation_sec} summarizes how \Vlasiator\ simulates the combined magnetosphere-ionosphere system, and describes the particular run analyzed in this study. Sections~\ref{fte_sec} and \ref{geoelectric_field_sec} respectively explain how we post-process the simulation data to find FTEs and calculate the geoelectric field.  Our results are presented in Section~\ref{results_sec}, highlighting the relationships between the simulated FTEs and geoelectric field signatures. We discuss the implications of our work in Section~\ref{discussion_sec} and draw conclusions in Section~\ref{summary_sec}.

\section{Simulation Description}\label{simulation_sec}

The \Vlasiator\ simulation presented here is the same as in the FTE study \cite{pfau-kempf_2025_fte_evolution_vlasiator}, and consists of a 3-dimensional hybrid-Vlasov     
model of the global magnetosphere \cite{palmroth25_vlasov_methods_living_review, ganse23} coupled to an electrostatic ionosphere model \cite{ganse_2025_vlasiator_ionosphere}.  The coupling scheme passes information between the magnetospheric and ionospheric spatial meshes, along dipolar field lines that are prescribed to fill the intervening region (idealized dipole with no tilt, moment $\bvec{m_E}=-8 \times 10^{22}\ \bvec{\hat{z}}\ \mathrm{m}^2\ \mathrm{A}$). Unless otherwise specified, spatial coordinates are given in the Geocentric Solar Ecliptic (GSE) coordinate system.

In the magnetospheric part of the simulation, the 3-dimensional proton velocity distribution function is evolved according to the Vlasov-Maxwell equations in the Darwin approximation, with the physics of the adiabatic background electron fluid entering through the pressure gradient and Hall terms of the generalized Ohm's Law. 
The simulation is held in a 3-dimensional Cartesian spatial box with dimensions $x\in[-110.5,50.2]\ \mathrm{R_E}$, $y\in[-57.8,57.8]\ \mathrm{R_E}$, $z\in[-57.8,57.8]\ \mathrm{R_E}$, where $\mathrm{R_E} = 6371$~km is Earth's radius. An approximate sphere constructed from the cubic mesh elements, at a radius of $r_{m}=4.7~\mathrm{R_E}$, sets the magnetosphere's inner boundary. 
An adaptive mesh refinement (AMR) scheme is used to define a Cartesian grid with spatially-varying resolution $\Delta x$ of the cubic mesh ``cells'' \cite{ganse23}, where the finest resolution $\text{min}(\Delta x)=1000\ \mathrm{km}$ is applied in regions such as the magnetopause that are of high scientific interest. A constant Maxwellian solar wind flows from the $+x$ boundary with fixed density, velocity, and temperature: $n_{sw} = 10^6\ \mathrm{m}^{-3}$, $\bvec{v_{sw}} = -7.5 \times 10^5\ \bvec{\hat x}\ \mathrm{m\ s^{-1}}$, $T_{sw} = 5 \times 10^5\ \mathrm{K}$. A constant southward magnetic field $\bvec{B_{sw}}=-5\ \bvec{\hat z}\ \mathrm{nT}$ is applied at the $+x$ boundary, embedded in the inflowing solar wind, to promote dayside reconnection. Homogeneous Neumann (``outflow'') boundary conditions are applied at the other box sides.
The velocity-space resolution of the proton distribution is $4\times 10^4$ m/s. The phase-space density threshold, below which the velocity distribution is neither stored nor propagated \cite{vonalfthan_2014_new_vlasiator_simulations, palmroth25_vlasov_methods_living_review}, is set to $10^{-15}\ \mathrm{m}^{-6}\ \mathrm{s}^3$ where the proton number density $n_p$ is $>10^5\ \mathrm{m}^{-3}$, to $10^{-17}\ \mathrm{m}^{-6}\ \mathrm{s}^3$ where $n_p <10^4\ \mathrm{m}^{-3}$, and linearly interpolated at intermediate $n_p$.

The ionosphere solver's input parameters are established using information from the magnetospheric domain. The field-aligned current is calculated for each ionospheric element by mapping along the field lines from a coupling radius $r_C\approx 5.6\ \mathrm{R_E}$ in the magnetosphere, assuming the total current is conserved along a flux tube. The ionospheric conductivity tensor $\mathbf{\Sigma}$ requires information about the incoming particles to determine the ionization rate. Therefore, the precipitating electron population density and temperature, precipitating energy flux, and field-aligned current density $J_\parallel$ are all sampled at $r_C$ and mapped to the ionospheric radius $r_B = R_E + 100~\mathrm{km}$ assuming collisionless dynamics along the field lines.
The downmapped quantities are temporally smoothed by an exponential filter  ($t_\text{smooth} = 4\,\text{s}$) to emulate the delay time of Alfv\'en waves propagating between the magnetosphere and ionosphere. 
Since the electron velocity distribution is not directly simulated in \Vlasiator, the ionospheric electron distribution is estimated with a proxy based on the ion distribution moments in the magnetosphere at radius $r_C$ \cite{ganse_2025_vlasiator_ionosphere}.  
Atmospheric ionization by solar photons is accounted for in the model, but the auroral field-aligned potential drop \cite{knight73} is neglected.

The simulation ran for 1612~$\mathrm{s}$, of which the first 500 seconds may be considered the initialization 
phase during which the magnetosphere forms and establishes a near-steady state \cite{ganse_2025_vlasiator_ionosphere}. The combined magnetosphere-ionosphere simulation state was saved every 1~second. This time resolution is sufficient to resolve the ground magnetic fluctuations relevant to GICs \cite{pulkkinen06}.

\section{Methodology}\label{methods_sec}

\subsection{FTE Identification}\label{fte_sec}

The relatively well-constrained 2D geometry of the magnetopause current sheet is conducive to the formation of ``quasi-2D'' structures during reconnection.
In particular, the FTEs formed by the magnetopause reconnection possess a characteristic 2D O-point topology, where magnetic field lines spiral around a central axis. Such spiral-field structures are more commonly referred to as flux ropes. The O-lines running through the flux ropes serve as proxies for the FTE locations.
Along with such ``O-lines'', analogous ``X-line'' structures are also formed, where the magnetic field in the plane perpendicular to a curvilinear axis exhibits an X-point topology, that is well-known from classical studies of 2D magnetic reconnection.  As a consequence, FTEs may be identified in 3D global or local simulations by detecting topological features where the magnetic field in the plane perpendicular to some axis vanishes.

We apply such a method for finding these X- and O- magnetic ``null lines'', as described in \citeA{alho24} in the context of a 3-dimensional magnetospheric \Vlasiator\ simulation. 
The ``O''-type magnetic topology at the center of an FTE can be readily identified in the LMN coordinate system, where the FTE axis is approximately along the $\bvec{\hat M}$-direction. We align our axes so that the current density along this direction is positive: $\bvec{J} \cdot \bvec{\hat M}>0$. A magnetic null line (where $B_L = B_N = 0$) may be classified as an O-line in the case $\pa B_N/\pa L < 0$, whereas for an X-line $\pa B_N/\pa L > 0$. As in \citeA{alho24}, we construct the orthogonal LMN coordinate system using a hybrid of the minimum directional derivative (MDD) and minimum gradient analysis (MGA) methods \cite<reviewed in >{shi19}. This combination avoids ambiguities in specifying unique LMN axes, which may otherwise arise when using the MDD or MGA method in isolation.

Of course, the field components $B_L, B_N$ evaluated at the discrete simulation locations (cell centers) are never exactly zero. So in practice, we identify null lines by expanding the field $\bvec{B}(\bvec{x})$ in a first-order Taylor series around each cell center $\bvec{x_0}$, i.e. assuming 
$\bvec{B}(\bvec{x_0 + \bvec{x^\prime}})\approx \bvec{B}(\bvec{x_0}) +  \nabla \bvec{B}(\bvec{x_0})\cdot \bvec{x}^\prime$, 
as described in \citeA{alho24}, following \citeA{fu_2015_FOTE_magnetic_nulls_MMS}. This results in a unique plane satisfying $B_L=0$, and another such plane satisfying $B_N=0$; their intersection yields a line along which $B_L = B_N = 0$. We identify the cell if this projected null line crosses through the cell's boundaries (the approximate region where the linear interpolation of $\bvec{B}(\bvec{x})$ is valid). We check for this intersection between the projected null line and cell surface using a signed distance function, a technique widely used in computer graphics. To accommodate ``near misses'', we identify a cell if the projected null line lies within a distance $0.2~\Delta x$ of any of the cell's surfaces. 
Once a null line is identified, it is classified as a X-line or O-line according to the sign of $\pa B_N/\pa L$ at that cell, as described above.

\begin{figure}
    \hspace{.5cm}
    \includegraphics[width=1.0\textwidth]{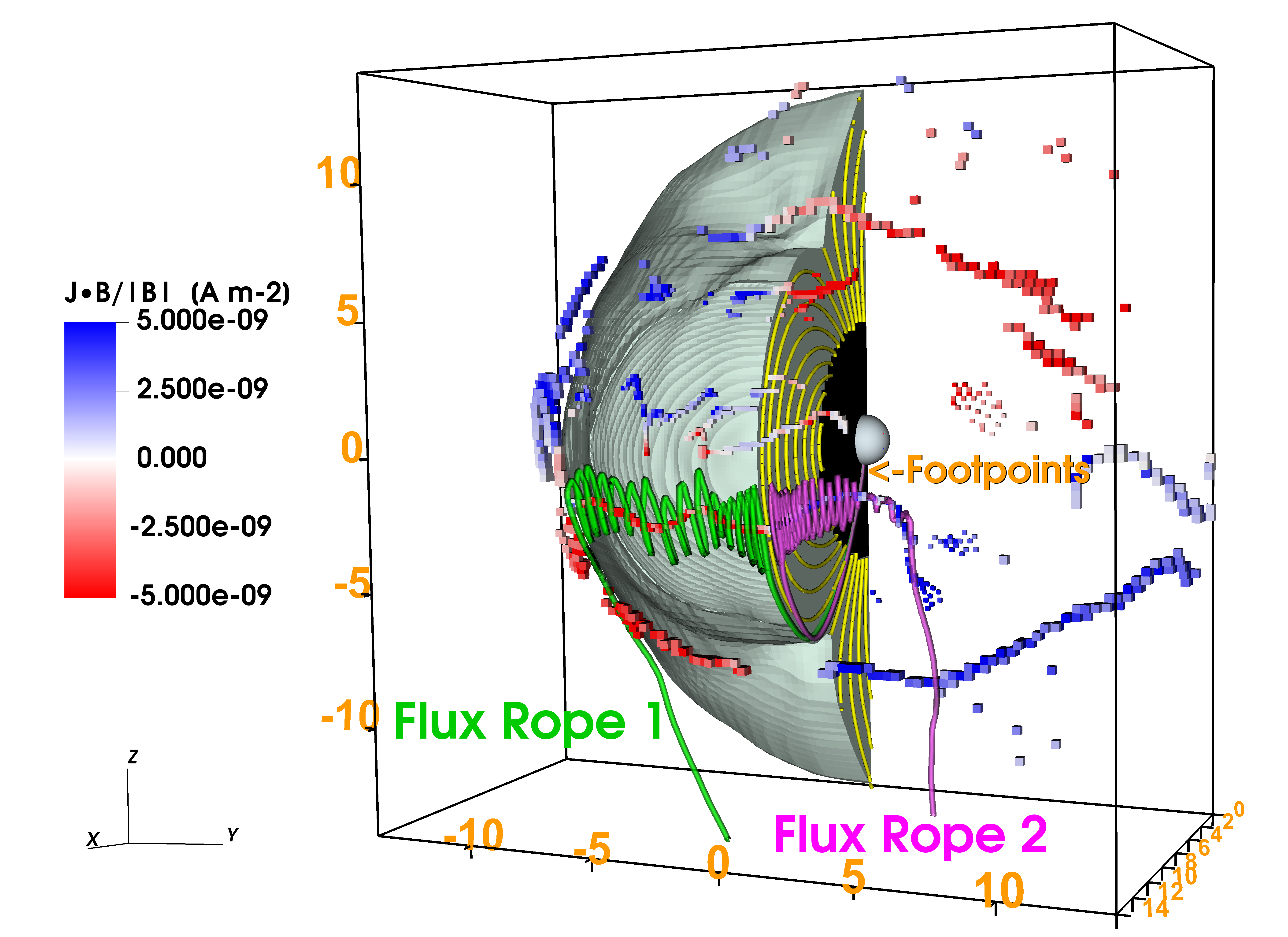}
    \caption{The FTEs at simulation time $t=1165\ \mathrm{s}$. The cutaway shows the $y<0$ half of the magnetopause, i.e. the $\beta^\star$=0.5 isosurface \cite{xu16, horaites23}. Yellow magnetospheric field lines are shown on the meridional plane for reference. \Vlasiator's ionosphere is shown by a central sphere, and the region $r_B<r<r_m$ between the magnetospheric and ionospheric domains is shown in black. The parallel current $J_\parallel$ evaluated at the FTE O-lines is shown in a blue-red scale. Two Flux Ropes, 1 and 2, in the same FTE are traced from starting positions $\bvec{x_1}$, $\bvec{x_2}$ on either side of a 3D magnetic null point (a blue-red junction). Both magnetic field lines are open-closed, with one end at the ionospheric footpoints of cusp field lines.}
    \label{fte_xo_fig}
\end{figure}

An example of the FTE identification for our \Vlasiator\ simulation is shown in Figure~\ref{fte_xo_fig}, at the time step $t=1165\ \mathrm{s}$. The O-lines, which appear as strands that are several $\mathrm{R_E}$ in length, are plotted in a subdomain of the simulation: $x\in [0,14]~\mathrm{R_E}$, $y\in [-14,14]~\mathrm{\mathrm{R_E}}$, $z\in [-14,14]~\mathrm{\mathrm{R_E}}$. The magnetopause is shown as a light green surface, which we define by the isocontour of the modified plasma beta, $\beta^\star = 0.5$ \cite{xu16, horaites23}. At the O-line cells, the field-parallel current density $J_\parallel = \bvec{J} \cdot \bvec{B} / |\bvec{B}| \approx \bvec{J}\cdot \bvec{\hat M}\ \mathrm{sign}(B_M)$ is shown in a red-blue color scale.

We observe that $J_\parallel$ commonly changes sign along the O-lines, at the red-blue junctions.
Because of the convention $\bvec{J} \cdot \bvec{\hat M}>0$ we use to define the LMN coordinate system, the $J_\parallel$ sign reversal must be explained by a corresponding sign reversal in the axial $B_M$ component of the O-line's magnetic field. 
Since the off-axis components $B_L, B_N$ are by definition also zero on a null line, we identify the red-blue junctions in Fig.~\ref{fte_xo_fig} as {\it 3-dimensional (3D) magnetic nulls} ($\bvec{B}=0$).

Each of these flux ropes' field lines is rooted on one end in the ionosphere, while the other end extends into the solar wind. To illustrate this, in Fig.~\ref{fte_xo_fig} we trace the magnetic field lines starting from two seed points, $\bvec{x_1} = [10.45, -1.43, -2.6]~\mathrm{R_E}$ and $\bvec{x_2} = [10.5, 1.47, -2.56]~\mathrm{R_E}$, that are on opposite sides of a null point. 
The traces form two distinct magnetic coils that wrap around the O-line, which we label as ``Flux Rope 1'' and ``Flux Rope 2''. We observe that the traced field lines in Fig.~\ref{fte_xo_fig} map down to Earth's ionosphere along the southern magnetic cusp, as may be expected because the null point separating the two flux ropes is near the noon meridian. Writing the angular coordinates on the spherical ionospheric mesh as an ordered pair [$\phi$, $\lambda$] of longitude ($\phi$) and latitude ($\lambda$), we find that $\bvec{x_1}$ maps to the coordinates $P_1 = [-2.07^\circ, -78.75^\circ]$ and $\bvec{x_2}$ maps to $P_2 = [2.48^\circ, -78.75^\circ]$. That is, the magnetic footpoints of Flux Ropes 1 and 2 are spaced around the noon meridian at a latitude of $\sim$79$^\circ$ South.

We evaluate the orientation of the magnetic field twist around the O-lines' axial field using $\text{sign}(J_\parallel)$, a simple measure of the ``current helicity''  that is commonly used in studies of the Sun \cite{van_driel-gesztelyi_2003_solar_helicity_observations}. The current helicity is technically distinct from the ``magnetic helicity'', another measure of twistedness, but these can often be assumed to have the same sign for flux ropes found in space \cite{russell_2019_current_magnetic_helicities_same_sign}. In Fig.~\ref{fte_xo_fig}, the central O-line of Flux Rope 1 has left-handed ``LH'' helicity ($J_\parallel <0$), while Flux Rope 2 has right-handed ``RH'' helicity ($J_\parallel >0$).  In supplementary Figure~S1, we trace additional FTE field lines for the time step shown in Fig.~\ref{fte_xo_fig}. A clear trend can be seen, where several FTEs are bifurcated around magnetic null points (i.e. the red blue current junctions), forming flux rope pairs with opposite helicity.

In Movie S2, we animate the evolution of Fig.~\ref{fte_xo_fig} throughout the simulation, showing $J_\parallel$ in the same red-blue coloring. The FTEs are generated near the equator and execute a complex writhing motion in addition to their expected poleward propagation \cite<e.g., >[]{hoilijoki_2019_FTE_2d_vlasiator, pfau-kempf_2020_dayside_3d_reconnection}---complicated by the merging and splitting of existing O-lines. The FTEs are typically observed to disintegrate shortly after arriving at the magnetic cusps (which appear as dimples in the $\beta^\star$ magnetopause surface), an indication of cusp reconnection \cite{omidi_2007_cusp_reconnection, paul23, pfau-kempf_2025_fte_evolution_vlasiator}.

\subsection{Geoelectric Field}\label{geoelectric_field_sec}

In the so-called ``plane-wave model'', the geoelectric field $\bvec{E}(t)$ is induced by magnetic fluctuations of an electromagnetic wave propagating vertically towards Earth's surface \cite{pirjola02}. Decomposing this horizontal ground electromagnetic field at a given position into its orthogonal northern (latitude $\lambda$) and eastern (longitude $\phi$) components, we may write the geoelectric field as an integral over the time history of the surface magnetic field: 

\begin{equation}\label{cagniard_eq}
E_{\phi}(t) = \frac{1}{\sqrt{\pi \mu \sigma}}\int_0^\infty \frac{d B_{\lambda}(t-t^\prime)}{dt}\frac{1}{\sqrt{t^\prime}} dt^\prime 
\end{equation}

\begin{equation}\label{cagniard_eq2}
E_{\lambda}(t) = \frac{-1}{\sqrt{\pi \mu \sigma}}\int_0^\infty \frac{d B_{\phi}(t-t^\prime)}{dt}\frac{1}{\sqrt{t^\prime}} dt^\prime,
\end{equation}

\noindent where the constants $\mu$, $\sigma$ are respectively the permeability and conductivity of the conducting crust. Following \citeA{pulkkinen06}, we adopt the typical values $\mu = 4 \pi \times 10^{-7}\ H/m$ and $\sigma=10^{-3}\ S/m$, while noting that in practice the surface conductivity can vary by multiple orders of magnitude across geographic locations \cite<e.g., >[]{marshalko_2021_geoelectric_field_3_approaches}. Equations (\ref{cagniard_eq}), (\ref{cagniard_eq2}) were first reported in \citeA{cagniard53}, and a detailed derivation can be found in \citeA{love14}.

To calculate the induced geoelectric field with equations~(\ref{cagniard_eq}) and  (\ref{cagniard_eq2}), 
we must first determine the horizontal components of the fluctuating magnetic field, $\bvec{B}(\bvec{x}, t)$, for each location $\bvec{x}$ at Earth's surface. We note that the ground magnetic field is a superposition of the fields produced by ``internal'' and ``external'' source currents, which each contribute significantly \cite{juusola_2020_internal_vs_external_B_field}: $\bvec{B}(\bvec{x}, t) = \bvec{B^{int}}(\bvec{x}, t) + \bvec{B^{ext}}(\bvec{x}, t)$. The external sources are above the Earth's surface (horizontal ionospheric currents, FACs, etc.), while the internal sources are below the surface (telluric currents).  
To estimate $\bvec{B^{int}}$, we will apply the coarse approximation that the Earth's surface behaves like an infinite, perfectly-conducting plane \cite<in a similar spirit as>{pirjola_1989_perfect_conductor_below_surface, pulkkinen_2003_SECS_Bint_Bext_separation}.
In this case, the horizontal components ($\phi$ and $\lambda$) of $\bvec{B^{ext}}$ are doubled by the action of the internal surface currents, i.e. $B^{int}_{\{\phi,\lambda\}} = B^{ext}_{\{\phi,\lambda\}}$, or equivalently:

\begin{equation}\label{Btot_2Bext_eq}
B_{\{\phi, \lambda\}}(\bvec{x}, t) = 2\ B^{ext}_{\{\phi, \lambda\}}(\bvec{x}, t).
\end{equation}

\noindent Equation (\ref{Btot_2Bext_eq}) may be applied to the finite-$\sigma$ case, when the horizontal spatial scale of $\bvec{B^{ext}}(\bvec{x}, t)$ is much larger than the frequency-dependent skin depth $\delta(f)=(\pi \mu \sigma f)^{-1/2}$; see equation 6 of \citeA{schmucker_1970_induction_anomalies} in the ``Tikhonov-Cagniard approximation'' ($k = 0$). In this paper, we will assume that the dominant magnetic field fluctuations occur at frequencies $f \gtrsim 0.01 \ \mathrm{Hz}$, so that Equation (\ref{Btot_2Bext_eq}) above is valid for surface magnetic fields at horizontal spatial scales $\gg 160\ \mathrm{km}$. This is a reasonable assumption for our simulation, which has a minimum ionospheric grid resolution of $l_{\mathrm{eff}}\sim100\ \mathrm{km}$ \cite{ganse_2025_vlasiator_ionosphere}.
In addition to the 1D horizontal spatial profiles analyzed in \citeA{schmucker_1970_induction_anomalies}, the ratio between $\bvec{B^{ext}}$ and $\bvec{B^{int}}$ has been quantified in equations~6-7 of \citeA{kuvshinov_2008_3D_global_geomagnetic_induction_model} in a more realistic 3D geomagnetic model based on spherical harmonics and a 1D radial conductivity profile $\sigma(r)$---they also find that the perfect conductor assumption (\ref{Btot_2Bext_eq}) is a good approximation for applied signals with periods $1/f\lesssim 2\ \mathrm{min}$.

Following \citeA{shao02, welling20}, we compute $\bvec{B^{ext}}(\bvec{x}, t)$ at each element of the ionospheric mesh according to the Biot-Savart law, an integral over the current density $\bvec{J}(\bvec{x}, t)$:

\begin{equation}\label{biot_savart_eq}
{\bf B^{ext}}({\bf x}, t) = \frac{\mu_0}{4\pi} \int_{\mathcal{V}} \frac{ {\bf J}({\bf x} - {\bf x^\prime}, t) \times {\bf x^\prime} }{|{\bf x}^{\prime}|^3} d{\bf x^\prime}.
\end{equation}

\noindent To compute the discrete approximation of eq.~(\ref{biot_savart_eq}), we divide the simulation volume $\mathcal{V}$ into 3 Zones, distinguished by the radial distance $r$:

\begin{itemize}
    \item Zone 1 (Magnetosphere): $r \geq r_C$\\
    \item Zone 2 (FACs): $r_C > r > r_B$\\
    \item Zone 3 (Ionosphere): $r = r_B$
\end{itemize}

\noindent The contributions of Zones~1 and~3 to the integral (\ref{biot_savart_eq}) are calculated directly from the \Vlasiator\ data. The Zone~3 integral is technically an analogous surface integral over the ionospheric mesh, where $\bvec{J}(\bvec{x}, t)$ is replaced by the height-integrated current density $\bvec{j}(\bvec{x}, t)$ that is determined by the ionospheric solver \cite<units $\mathrm{A\ m^{-1}}$, see>{ganse_2025_vlasiator_ionosphere}. The Zone~2 integral requires a prescription for the current density because the simulations do not represent this region directly. In Zone~2 we assume that along each field line the current density is field-aligned and divergenceless, i.e. $\bvec{J}(\bvec{x}, t) \propto \bvec{B}(\bvec{x}, t)$, in accordance with \Vlasiator's runtime coupling scheme (section~\ref{simulation_sec}). The magnetic field $
\bvec{B}$ in Zone~2 is assumed to be exactly dipolar, matching the magnetospheric domain's inner boundary condition at $r=r_m$. The numerical approximation of the Zone~2 integral requires higher resolution than for Zone~1, because the FAC features at the coupling radius (i.e., at $r=r_C$) are focused by the magnetic field geometry to a finer scale in Zone~2.

We numerically calculate the integrals (\ref{cagniard_eq}), (\ref{cagniard_eq2}) over the finite duration of the simulation.  Prescriptions for these numerical integrals, which must handle the singularity at $t^\prime =0$, are given in \citeA{liu09, love14}. We approach the integration by treating $B_\lambda(t)$ and $B_\phi(t)$ as piecewise linear functions that interpolate the discrete time series data, so that $dB_\lambda/dt$ and $dB_\phi/dt$ are piecewise constant. In this case, the $1/\sqrt{t^\prime}$ dependence in 
(\ref{cagniard_eq}), (\ref{cagniard_eq2}) is integrable, with the exact analytical expression given by Eq.~5 of \citeA{liu09}. To evaluate the integrals, we ignore the simulation's initialization phase and therefore set $d \bvec{B}/ d t = 0$ during the times $t \leq500\ \mathrm{s}$. With this approach, we calculate the geoelectric field at 1~$\mathrm{s}$ resolution for every element in the ionospheric mesh.

\begin{figure}
    \hspace{.5cm}
    \includegraphics[width=1.0\textwidth]{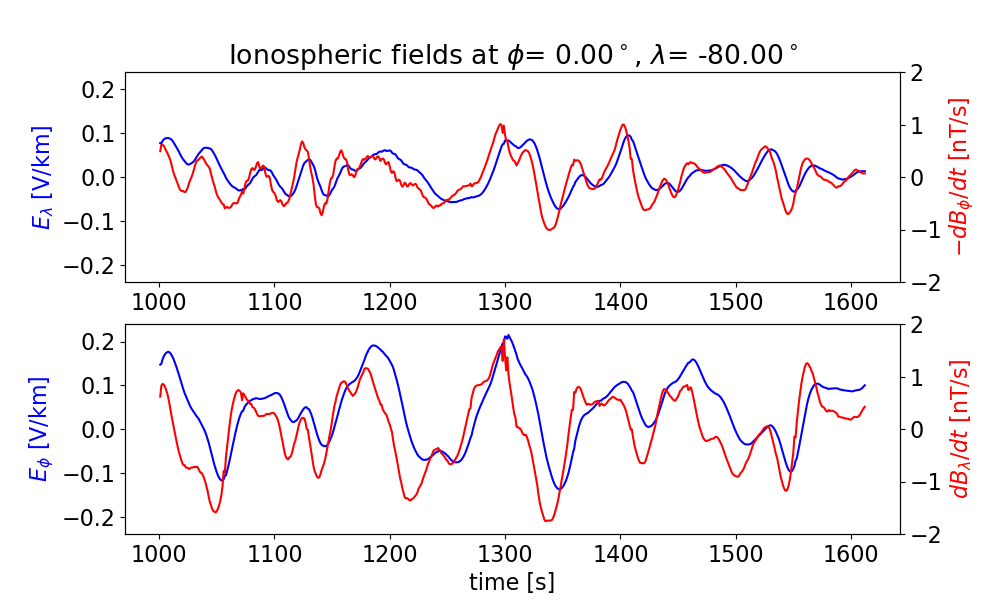}
    \caption{Ground electromagnetic fields near the footpoints of cusp fieldlines in the southern hemisphere.  The northern and eastern geoelectric field components $E_\lambda$, $E_\phi$, showing also their respective correlations with $-dB_\phi/dt$, $dB_\lambda/dt$.}    
    \label{Exy_dByx_fig}
\end{figure}

Figure~\ref{Exy_dByx_fig} shows the geoelectric field components $E_\lambda$, $E_\phi$ evaluated at a point [$\phi$, $\lambda$] = [$0^\circ$, $-80^\circ$] located $\sim$1$^\circ$ south of the footpoints of Flux Ropes 1 and 2 pictured in Fig.~\ref{fte_xo_fig}. The plot reproduces the well-known correlations between $E_\lambda$ and $-dB_\phi/dt$, as well as the analogous correlation between $E_\phi$ and $dB_\lambda/dt$ \cite<e.g., >{pulkkinen06}. For this position, the eastward electric field $E_\phi$ has a peak amplitude of $\sim$0.2~V/km, approximately double the amplitude of the northward $E_\lambda$ component. These fields, which we observe under moderate solar wind driving conditions, are weaker than but still within an order of magnitude of the $|\bvec{E}|=1\ \mathrm{V/km}$ threshold that has been used to classify ``extreme'' GIC events \cite{pulkkinen_2015_extreme_geoelectric_field_statistics, ngwira_2015_extreme_geolectric_fields}.

\section{Analysis}\label{results_sec}

\subsection{Flux ropes and 3D nulls}\label{3d_nulls_sec}

The apparent 3D magnetic null point we observe, that separates Flux Ropes 1 and 2 in Figure~\ref{fte_xo_fig}, is significant because 3D magnetic nulls are known in the context of magnetic reconnection \cite{greene_1988_properties_of_3d_magnetic_nulls_reconnection, lau_finn_1990_3d_magnetic_nulls_reconnection}. These null points have several remarkable properties. 
First, 3D null points are connected by 2D (X- and O-) null lines, such as FTE O-lines in our study \cite<e.g., >{murphy_2015_appearance_and_disappearance_of_3d_magnetic_nulls}.
Furthermore, 3D null points appear in pairs with opposite ``polarity'' $\Pi$, which is defined by $\Pi = \text{sign}( \det(\nabla \bvec{B}))$ evaluated at the null point. Just as they are created, null pairs may annihilate in opposite-polarity pairs, so the sum of polarities among the null points is always zero.
Such null points have been observed in Earth's magnetosheath by the Cluster and MMS spacecraft \cite{wendel_adrian_2013_magnetic_nulls_cluster, fu_2018_magnetic_nulls_mms_observations}.

\begin{figure}
    \includegraphics[width=1.0\textwidth]{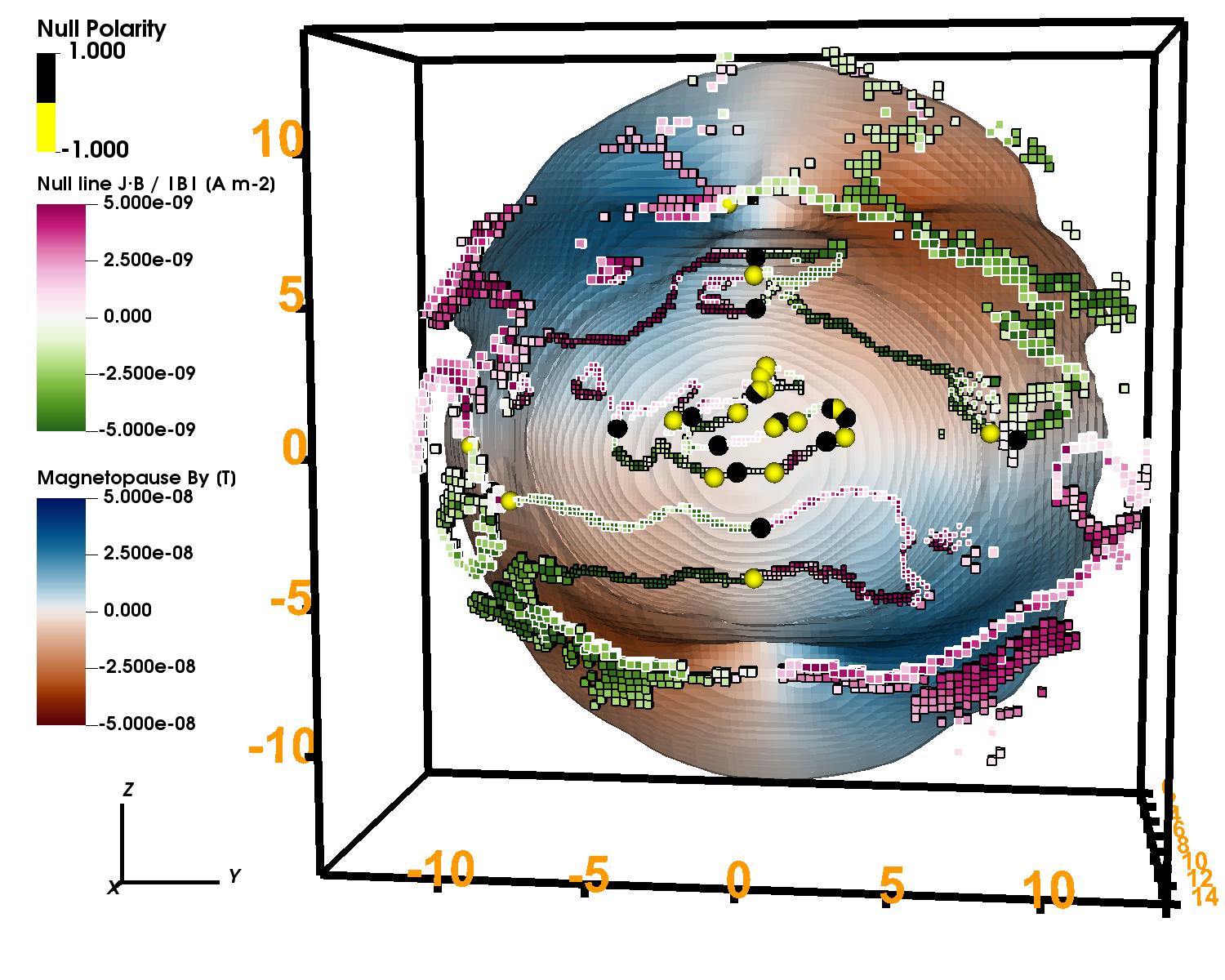}
    \caption{The $\beta^\star=0.5$ magnetopause is shown head-on, at $t=1165\ \mathrm{s}$ in a spatial box $x\in [0, 14~ \mathrm{R_E}]$, $y\in [-14~\mathrm{R_E}, 14~ \mathrm{R_E}]$, $z\in [-14~\mathrm{R_E}, 14~ \mathrm{R_E}]$. The magnetopause surface is colored in a red-blue scale according to the magnetic field component $B_y$. Simulation cells containing null lines are colored by the parallel current $J_\parallel$ (green-purple), and the cell edges are highlighted in white (for O-lines) or black (for X-lines). The 3D null points, which we observe occur on the null lines at locations where $J_\parallel$ changes sign, are shown as spheres. The null points are colored based on their polarity: yellow if $\Pi=-1$ and black if $\Pi=1$. This figure is animated in movie~S3.
    }
    \label{fte_data_fig}
\end{figure}

In Figure \ref{fte_data_fig}, we show the relationships between the X- and O-lines, 3D magnetic nulls, and the magnetopause geometry. At the same time step pictured in Fig.~\ref{fte_xo_fig}, $t=1165\ \mathrm{s}$, we now show both the X- and O-lines---the edges of the X-line cells are highlighted in black, whereas the O-lines are highlighted in white. The current density along the null lines, $J_\parallel$, is shown in a green-purple scale. 
We observe $\text{sign}(J_\parallel)$ at the null lines tends to flip between north-south and east-west hemispheres, so that $\text{sign}(J_\parallel)$ along the null lines is divided into four quadrants.
This same 4-quadrant structure is also observed in the $y-$component of the magnetic field, $B_y$, evaluated at the magnetopause surface (red-blue color scale).

We manually checked that the current in the null lines is generally oriented in the same direction as the eastward magnetopause current, i.e. roughly in the $\bvec{+\hat y}$ direction.  That is, the current flows along the null lines in a consistent sense. So, the reason $\text{sign}(J_\parallel) = \text{sign}(\bvec{J\cdot\bvec{B}})$ changes along the null lines is apparently due to a change in the sign of the null lines' axial field component $B_M$ in the LMN coordinate system---for these equatorially-stretched null lines, $B_M$  is approximately the same as the Cartesian $B_y$ component.  We therefore infer that the null lines inherit $\text{sign}(B_M)$, which determines the null line helicity, from $\text{sign}(B_y)$ of the underlying large-scale magnetopause magnetic field pattern.

The 3D null points in Fig.~\ref{fte_data_fig}, which are marked by their polarity as yellow ($\Pi =-1$) or black ($\Pi=1$), are found by spatially interpolating $\bvec{B}(\bvec{x})$ using the VisIt visualization software \cite{visit}. We observe that the 3D nulls occur at locations where $J_\parallel$ changes sign along the null lines.

In Movie S3, we animate the evolution of Fig.~\ref{fte_data_fig} throughout the simulation.
The FTEs are often segmented by 3D magnetic null points, into 2 or more regions with differing $\text{sign}(J_\parallel)$. The magnetic null points are frequently found near the equator and the noon meridian. During the ongoing magnetopause reconnection, the 3D null point pairs (yellow-black) undergo a continual process of creation and annihilation.

\begin{figure}

    \hspace{2cm}\includegraphics[width=0.7\textwidth]{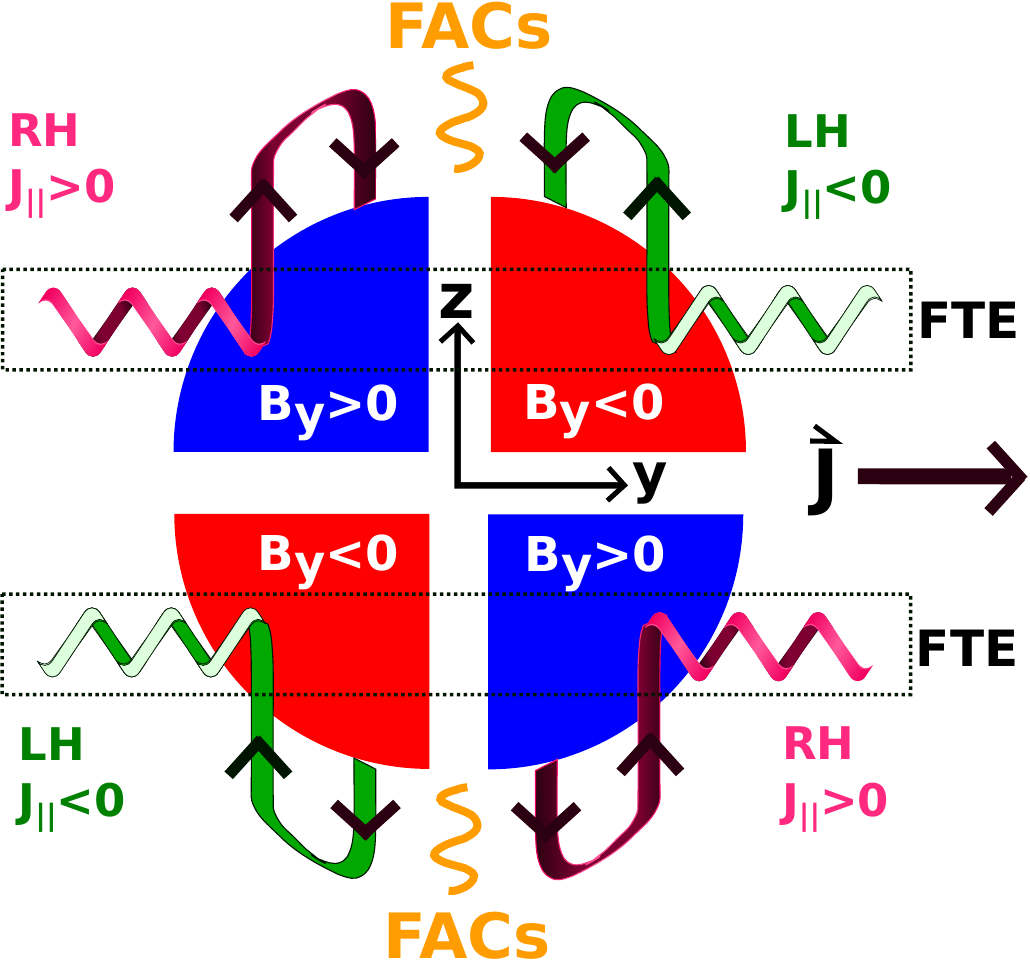}
    \caption{ Schematic diagram showing the magnetic field of the FTE flux ropes, as projected in the simulation's $y$-$z$ plane. The current helicity (RH or LH) and $\text{sign}(J_\parallel)$ of the flux ropes are correlated with $\text{sign}(B_y)$ of the underlying dayside magnetosphere. These quantities are thus organized into the four quadrants of the $y$-$z$ plane. The FTE current $\bvec{J}$ flows in the $+\bvec{\hat y}$ direction, in agreement with the magnetopause current. FACs flow to Earth's ionosphere, to a region containing the footpoints of the dayside cusp field lines. 
    }
    \label{fte_schematic_fig}
\end{figure}

We illustrate these various trends in Figure~\ref{fte_schematic_fig}, in a y-z plane projection of the magnetopause-FTE system. The figure shows an FTE in both the northern and southern hemispheres, each of which is bisected near the meridian ($y=0$) into two flux ropes with opposite $\text{sign}(J_\parallel)$ and helicity (RH or LH). A 3D magnetic null (not shown) is present at the junction of each flux rope. For comparison, we show schematically $\text{sign}(B_y)$ of the underlying $\beta^\star=0.5$ magnetopause---i.e. on the magnetospheric side of the magnetopause current sheet. In our simulation, $\text{sign}(B_y)$ on this surface exhibits a 4-quadrant structure when projected in the y-z plane, which is the same pattern of $\text{sign}(B_y)$ created by Earth's magnetic dipole field.  The transmission of transient FAC signals along the cusp field lines, a topic we address in section~\ref{facs_sec}, is shown schematically.

\subsection{FAC Transmission}\label{facs_sec}

Here we provide evidence that the coupling between Earth's ionosphere and FTEs in the magnetosphere is mediated by field-aligned currents. To demonstrate this link, we use our method of FTE identification (section~\ref{fte_sec}) in combination with tracking the FAC propagation. In \Vlasiator, each magnetosphere-coupled element in the ionospheric mesh maps to a cell located at the coupling radius $r_C$, and this mapping is used to input FAC information into the ionospheric solver (section~\ref{simulation_sec}).  Therefore, to complete the chain connecting the ionosphere and the FTEs, we must track how the FACs flow from the magnetospheric FTEs (at $r>r_C$) down to the coupling radius $r_C$.

For this purpose, we first trace a magnetic field line at time $t_0=1250\ \mathrm{s}$, forming a curve $C_{34}$ that extends a curvilinear distance $d_C = 7.85~\mathrm{R_E}$ between the positions $\bvec{x_3}$=[2.75, 0.60, 4.88]~$\mathrm{R_E}$ and $\bvec{x_4}$=[8.90, 2.77, 4.34]~$\mathrm{R_E}$. Secondly, we define the radial segment $C_{45}$ by tracing radially outwards for an additional distance $d_R = 2~\mathrm{R_E}$, from $\bvec{x_4}$ to the point $\bvec{x_5}$=[10.67, 3.32, 5.20]~$\mathrm{R_E}$. The final curve $C_{345}$ is given by their union: $C_{345} = C_{34} \cup C_{45}$. This curve lies entirely in the quadrant $y>0$, $z>0$. We remark that the position $\bvec{x_4}$ is just inside the magnetopause, so that FTEs pass near to this location as the simulation progresses. Indeed, the radial segment $C_{45}$ is designed to intersect the O-lines of the FTEs as they convect around the magnetopause. The passage of these FTEs may be expected to disturb the plasma near $\bvec{x_4}$, launching parallel-propagating waves along $C_{34}$ towards $\bvec{x_3}$. Any FACs reaching $\bvec{x_3}$ (which is at the coupling radius $r_C$) are transmitted to the ionosphere along the magnetic field lines, i.e. to the mesh element at coordinates $[\phi, \lambda]$=$[12.1^\circ, 78.0^\circ]$. 

\begin{figure}
    \includegraphics[width=1.0\textwidth]{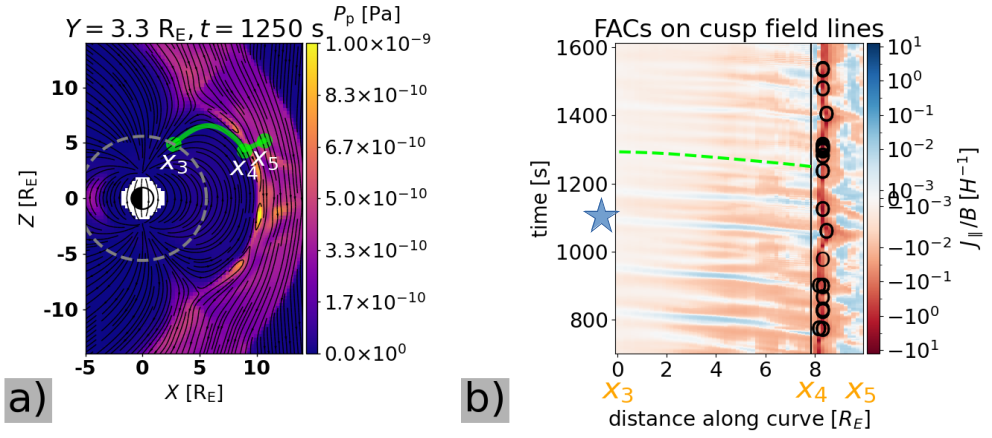}
    \caption{{\bf a)} Proton plasma pressure $P_p$ in the plane $y=3.3~\mathrm{R_E}$, at time $t_0=1250$~s. The FTEs appear as regions of enhanced plasma pressure at the magnetopause, encircled by magnetic field lines (black). The projection of the curve $C_{345}$ in the plane is shown as a green line, with the points $\bvec{x_3}$, $\bvec{x_4}$, $\bvec{x_5}$ shown as green dots. The coupling radius $r_C$ is shown as a grey dashed circle. {\bf b)} The scaled parallel current $J_\parallel/B$, shown in a time elongation map where the x-axis represents the distance along the curve $C_{345}$ and the y-axis represents time. Each occasion that an FTE O-line crosses the curve $C_{345}$ is marked with an ``O''. The green dashed line, shown for reference, represents the space-time propagation of a field-aligned Alfv\'en wave, moving along $C_{345}$ from $\bvec{x_4}$ to $\bvec{x_3}$ and starting at initial time $t_0$. A blue star marks the arrival of an FAC pulse at the inner boundary at $t\approx1100\ \mathrm{s}$.}
    \label{facs_propagation_fig}
\end{figure}

To illustrate, a representative planar cut ($y=3.3 \mathrm{R_E}$) of the \Vlasiator\ magnetosphere at time $t_0$ is shown in Figure~\ref{facs_propagation_fig}a.
In addition to the color map, which shows the proton plasma pressure, the projection of the curve $C_{345}$ into the plane is shown as a green line. 
The magnetic fields in the inner magnetosphere are relatively steady, in part due to the simulation's constant driving conditions, so we may adopt the subsegment $C_{34}$ as a reasonable proxy for the path of the FACs throughout the simulation duration. Of course, the accuracy of this approximation is correlated with proximity to Earth, and in particular we acknowledge that near the magnetopause the ongoing reconnection is expected to lead to complicated time-evolution of the magnetic field.

In Figure~\ref{facs_propagation_fig}b, we show a time-elongation map (also known as a keogram or J-map), that demonstrates the propagation of field-aligned currents in space and time. 
The horizontal axis shows the distance along the fixed curve $C_{345}$.
Because of the expected scaling $J_\parallel\propto B$, in the time-elongation map we plot the normalized quantity $J_\parallel/B$ with a blue-red color scale.  We mark each occurrence of an O-line on the curve $C_{345}$, identified according to the criteria described in section~\ref{fte_sec}, with an ``O'' in the  Figure~\ref{facs_propagation_fig}b.
For reference, we sample the Alfv\'en speed $v_A$ along the curve $C_{345}$ at the time $t_0$, and use this to construct the space-time curve of a field-aligned Alfv\'enic pulse moving from $\bvec{x_4}$ to $\bvec{x_3}$ (green dashed line). From the figure, it is clear that enhanced Alfv\'enic signals are transmitted Earthward when O-lines cross the curve $C_{345}$ near the point $\bvec{x_4}$. The sign  $J_\parallel/B > 0$ indicates a field-aligned current pulse, which has the opposite sense of the parallel current of the FTEs. That is, $\text{sign}(J_\parallel/B)$ is negative at the FTE O-lines, as expected for positions $y>0$ in the northern hemisphere (see Fig.~\ref{fte_schematic_fig}).

\begin{figure}
    \includegraphics[width=1.0\textwidth]{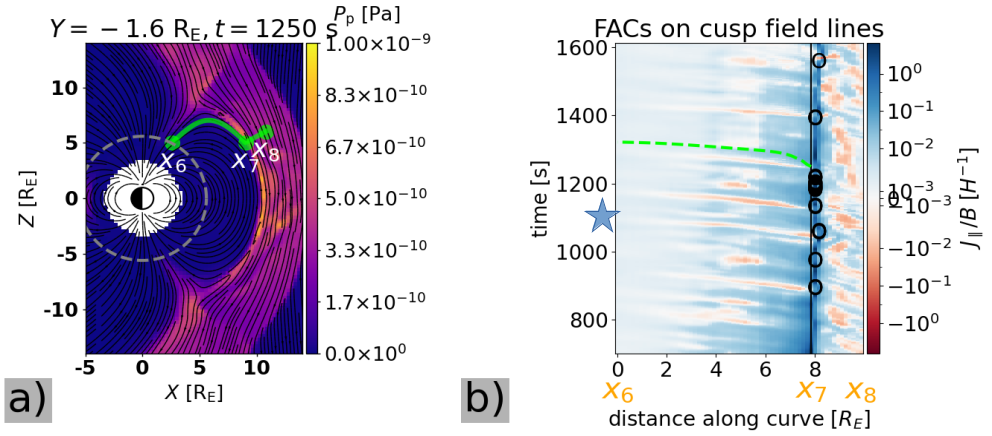}
\caption{{\bf a)} The projection of the curve $C_{678}$ overlaid on a plane cut ($y=-1.6\ \mathrm{R_E}$) of the simulation, at time $t_0=1250$~s, analogous to Fig.~\ref{facs_propagation_fig}a. {\bf b)} A time elongation map, analogous to Fig.~\ref{facs_propagation_fig}b, of the scaled parallel current $J_\parallel/B$ propagating from $\bvec{x_7}$ to $\bvec{x_6}$ when FTE O-lines cross the curve $C_{678}$. A blue star marks the arrival of an FAC pulse at the inner boundary at $t\approx1100\ \mathrm{s}$.}
    \label{facs_propagation_fig2}
\end{figure}

We repeat the preceding analysis for a curve $C_{678}$ that passes through an analogous set of points $\bvec{x_6}= [2.62, -0.26,  5.01]~\mathrm{R_E}$, $\bvec{x_7} = [9.02, -1.35,  5.04]~\mathrm{R_E}$, $\bvec{x_8} = [10.79, -1.61,  6.02]~\mathrm{R_E}$.  This curve lies in the quadrant $y<0$, $z>0$, i.e. on the opposite side of the noon meridian from $C_{345}$, and is near enough to intersect most of the same FTEs. Figure~\ref{facs_propagation_fig2}a shows the projection of this curve in the plane $y=-1.6\ \mathrm{R_E}$, with the proton pressure $P_p$ shown in a color scale, analogously to Fig.~\ref{facs_propagation_fig}a. Similarly, Figure~\ref{facs_propagation_fig2}b is analogous to Figure~\ref{facs_propagation_fig}b. Again we see a tendency for field-aligned current pulses to be launched along the field line from $\bvec{x_7}$ to $\bvec{x_6}$ when an FTE O-line passes near $x_7$. At time $t_0$, the location $\bvec{x_6}$ downmaps to $[\phi, \lambda]$=$[-5.7^\circ, 78.8^\circ]$. The propagation of the current pulses (red curves) again matches well with a representative Alfv\'enic trajectory starting from time $t_0$, shown by the green dashed line.  Here $\text{sign}{(J_\parallel/B)} < 0$ is again in the opposite sense of the parallel current of the FTEs overhead. The principal difference between Figs.~\ref{facs_propagation_fig} and \ref{facs_propagation_fig2} appears in the sign of the normalized parallel current $J_\parallel/B$, which is reversed between the two plots.

\subsection{Global Geoelectric Field}\label{geoE_global_map_sec}

\begin{figure}
    \includegraphics[width=1.0\textwidth]{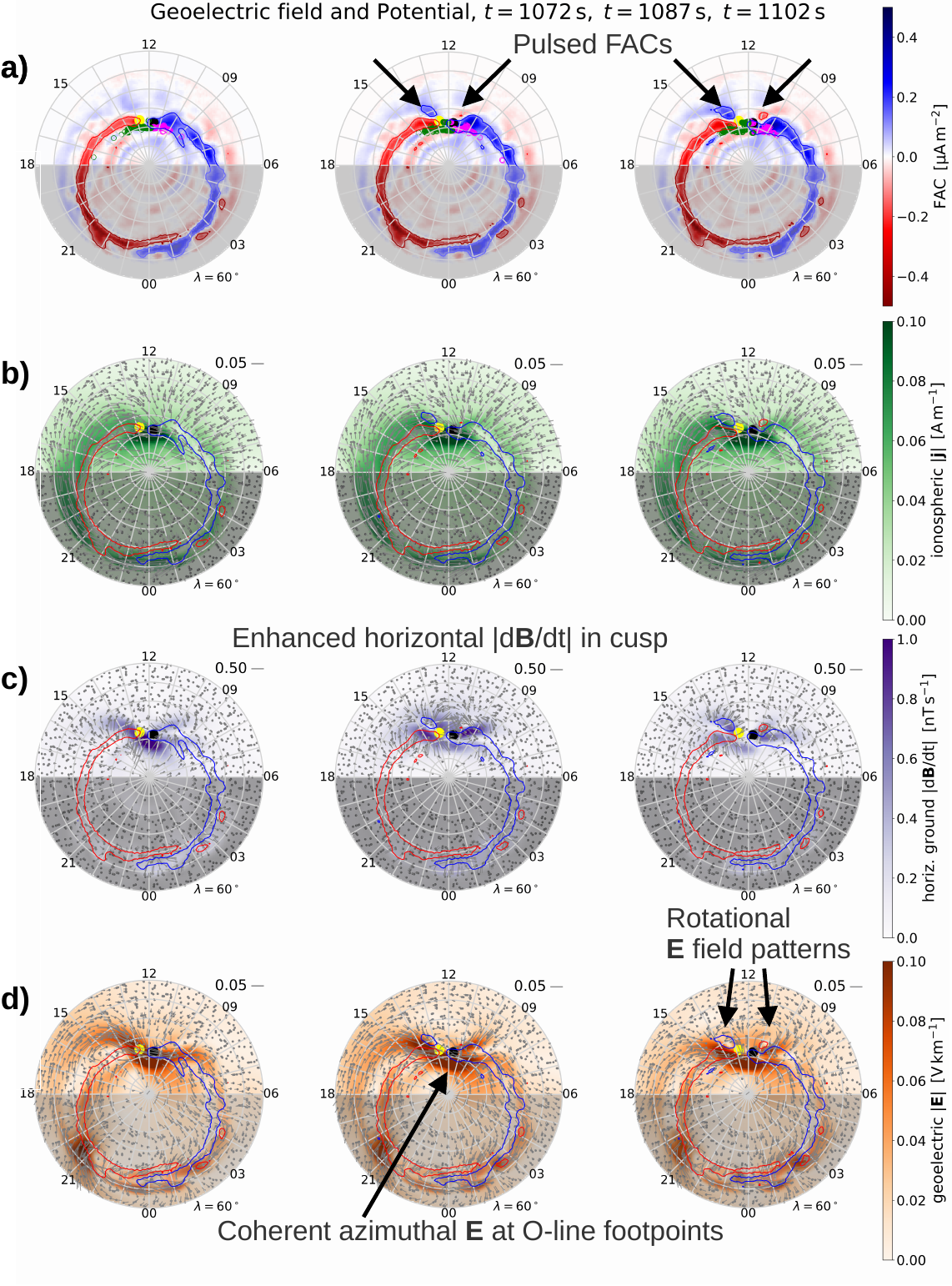}
    \hspace{0cm}
\caption{The geoelectric field and field-aligned currents in the northern polar cap ($\lambda>60^\circ$), at times $t=1072\ \mathrm{s}$, $t=1087\ \mathrm{s}$, $t=1102\ \mathrm{s}$ (left to right). Rows: {\bf a)} the field-aligned current (FAC) density $J_\parallel$ (color map), and the instantaneous footpoints of the dayside FTE O-lines are drawn as circles, colored by the O-line magnetic helicity---green (LH) or purple (RH), {\bf b)} the horizontal ionospheric current $\bvec{j}$ and its magnitude $|\bvec{j}|$ (color map), {\bf c)} the ground magnetic field time derivative $d\bvec{B}/dt$ and its magnitude $|d\bvec{B}/dt|$ (color map), {\bf d)} the geoelectric field $\bvec{E}$ and its magnitude $|\bvec{E}|$ (color map). The red-blue contours outlining the level set $|J_\parallel| = 0.1\ \mathrm{\mu A~m}^{-2}$ are shown in all plots. The magnetic footpoints of the magnetospheric locations $\bvec{x_3}$ and $\bvec{x_6}$ are respectively shown as yellow and black dots.} 
    \label{E_polar_cap_fig}
\end{figure}

We observe that the geoelectric field, calculated over Earth's surface as described in section~\ref{geoelectric_field_sec}, varies dynamically in the dayside magnetic cusp. Noting that the geoelectric field is largely driven by the horizontal ionospheric currents, which themselves are caused by FACs, we display both the fields and driving FAC currents. In Figure~\ref{E_polar_cap_fig}a-d, we show in separate rows
 the ionospheric field-aligned current density $J_\parallel$, the ionospheric height-integrated in-plane current density $\bvec{j}$, the time derivative of the horizontal ground magnetic field $d\bvec{B}/dt$, and the geoelectric field $\bvec{E}$.
These polar plots of the $\lambda > 60^\circ$ northern hemisphere show the magnetic local time ($\text{MLT} = 12 + \phi/15^\circ$) as the azimuthal component, while the grey latitudinal lines are spaced $\Delta \lambda = 5^\circ$ apart.  As our study focuses on the dayside cusp, we shade out the nightside data---recent \Vlasiator\ investigations of the magnetosphere-ionosphere coupling of FACs on the nightside are presented in \citeA{workayehu_2025_vlasiator_BBFs, koikkalainen_2025_vlasiator_nightside_M-I_coupling_PREPRINT}.

We consider the evolution of the cusp during the arrival of the pulsed magnetospheric field aligned currents. From left to right, the three columns of Figure \ref{E_polar_cap_fig} respectively show times $t=1072\ \mathrm{s}$, $t=1087\ \mathrm{s}$, $t=1102\ \mathrm{s}$ (15-$\mathrm{s}$ increments). This time range corresponds with the arrival of a current pulse at $\bvec{x_3}$ (marked with a star in Figure~\ref{facs_propagation_fig}b). Because the point $\bvec{x_3}$ is at the coupling radius $r_C$, the pulsed $J_\parallel>0$ current that arrives at $\bvec{x_3}$ maps along the field lines to the ionosphere. This time-dependent downmapped location is marked with a yellow dot in the plots, to show the approximate place where the pulsed FACs are expected to appear. In a similar fashion, we observe from Fig.~\ref{facs_propagation_fig2}b that a $J_\parallel<0$ current pulse should appear near the magnetic footpoint of $\bvec{x_6}$, in the western hemisphere---we mark this footpoint evaluated at the relevant time step with a black dot in each plot.

In Figure \ref{E_polar_cap_fig}a, we indeed see that over this time period, 1072-1102~$\mathrm{s}$, a pulse of positive current (blue, into the ionosphere) appears near the footpoint of $\bvec{x_3}$ (yellow dot) and a negative current (red, out of the ionosphere) appears near the footpoint of $\bvec{x_6}$ (black dot).  The level set of the ionospheric FACs, $|J_\parallel| = 0.1~\mathrm{\mu A~m^{-2}}$, outlines the Region~1 current system and is shown in all panels of the figure for reference (blue-red contours). Overlaid on the FAC plots, we also show the magnetic footpoints of the identified O-line cells in the dayside northern hemisphere ($x>0$, $z>0$). These footpoints are obtained by tracing the instantaneous magnetic field for a given time step in the magnetospheric domain of the simulation. The plotted circles marking these footpoint locations are colored based on the helicity of the field evaluated at the magnetospheric O-line cells, green (LH, $J_\parallel<0$) or purple (RH, $J_\parallel>0$) as before. We note that the plotted O-line footpoints are found slightly north of the Region 1 current system---overlapping with this region slightly. Where the O-line footpoints overlap with the Region~1 FACs, we observe that $J_\parallel$ on the magnetospheric O-line has the same sign as $J_\parallel$ at its footpoint (e.g., purple circles coincide with the blue Region~1 FACs). This indicates that the FACs in this region may be caused by the diversion of some of the FTE's axial current towards Earth.

Figures \ref{E_polar_cap_fig}b-c display quantities to provide context for observations, because in practice the geoelectric field is usually not measured directly but is rather inferred from the observed ground magnetic field fluctuations and the inferred ionospheric current. Figure \ref{E_polar_cap_fig}b shows the horizontal, height-integrated ionospheric current $\bvec{j}$. The magnitude $|\bvec{j}|$ is shown in a color map, while the vector components are displayed in a ``feather plot'' style.  The horizontal currents are perturbed by the arrival of the FAC pulse, but the deviation is minor, as these currents are dominated by the near-steady state pattern. Figure \ref{E_polar_cap_fig}c shows the horizontal components of the vector field $d\bvec{B}/dt$, and the magnitude $|d\bvec{B}/dt|$. We note that the $|d\bvec{B}/dt|$ is generally enhanced in the cusp and around the location of the time-varying FACs.

At the same time the FAC pulse arrives in the ionosphere, the geoelectric field $\bvec{E}$ also undergoes a dramatic change. As seen in Figure~\ref{E_polar_cap_fig}d, at time $t=1072\ \mathrm{s}$, the field $\bvec{E}$ points generally southward in the region MLT$\sim$12, $\lambda\sim75^\circ$. But after the pulsed FACs arrive, at $t=1102\ \mathrm{s}$ we observe that $\bvec{E}$ develops into a pair of rotational cells. Where the pulsed currents point into the ground (blue contour), the field $\bvec{E}$ rotates clockwise, and conversely $\bvec{E}$ rotates counterclockwise where the pulsed currents point out of the ground (red contour). Analogous behavior has been observed in the divergence-free component of the ionospheric current $\bvec{j}$, which likewise rotates around the FAC enhancements that are found at the center of TCVs \cite{amm_2002_tcv_event_study}. The strongest geoelectric field magnitudes occur slightly north of these rotational structures, at latitudes $\lambda \sim80^\circ$, coinciding with the footpoints of the FTE O-lines (locations marked in Fig.~\ref{E_polar_cap_fig}). In the time interval shown, $1072-1102\ \mathrm{s}$, this region begins as a coherent eastward geoelectric field stretching across a wide range of longitudes ($10<\mathrm{MLT}<14$). However, this directional coherence is perturbed by the arrival of the FAC pulse.

The complex time-evolution of the geoelectric field and field-aligned currents is shown in the supplementary movie S4. This movie animates the single-time plots shown in Figure~\ref{E_polar_cap_fig}. The animation shows many bursts of activity in the cusp region, where the dayside geoelectric field is most significant. As in Figure~\ref{E_polar_cap_fig}, we observe the TCV-like rotational patterns in the vector field $\bvec{E}$ coincide with the appearance of pulsed FACs near the noon meridian, and these disturbances appear to propagate along the auroral oval towards the nightside before eventually dissipating. Besides the pulsed FAC event described above at $t=1100\ \mathrm{s}$, the cusp region behaves similarly around times $t=650\ \mathrm{s}$ and $t=950\ \mathrm{s}$. The animation S4 also shows the evolution of the high-magnitude geoelectric field observed around $\lambda \sim80^\circ$, near the FTE O-line footpoints.

\section{Discussion}\label{discussion_sec}

\subsection{O-line Helicity and Magnetopause $B_y$}

Our observation that the FTE O-line helicity $\text{sign}(J_\parallel)$ correlates with the underlying magnetopause $B_y$ is similar to prior studies, which have found that the helicity is related to the solar wind IMF $B_y$.
For example, \citeA{leefu85} found that reconnection at the magnetopause can result in pairs of flux ropes, with opposite helicity, that propagate towards opposite poles. In this picture, the flux rope guide field is directly determined by the IMF $B_y$, which then sets the helicity of the reconnecting flux ropes.  This mechanism was confirmed in an observational study of FTEs, using MMS data \cite{kieokaew21}; however, in that study 1 of 6 events were termed ``outliers'' because they did not match the predicted pattern. These outlier events were addressed in another observational study \cite{dahani22}, which suggested that when the IMF $B_y$ is sufficiently weak, the FTE guide field can be generated by the Hall effect. The authors found that the outliers occurred in regions of the magnetopause with high magnetic shear (i.e. the quadrants $y<0, z>0$ and $y>0, z<0$). These Hall fields are known from simulations of asymmetric reconnection \cite{karimabadi99}. 

In our simulations, the IMF is purely southward and aligned with Earth's dipole moment, and the Hall effect is incorporated in the $\bvec{J}\times \bvec{B}$ term of the generalized Ohm's Law. Since the IMF $B_y$ is absent, we may expect the Hall effect to be significant during the dayside reconnection that produces FTEs. However, this effect, which is known from 2D reconnection simulations, predicts only a north-south hemispheric dependence of the flux rope helicity, and not the east-west dependence that we also observe (Fig.~\ref{fte_data_fig}). Rather, the four-quadrant pattern of the helicity $\text{sign}(J_\parallel)$ of our simulated flux ropes is clearly organized by the magnetopause $B_y$. Therefore, for our simulations we conclude that the flux rope guide field is predominantly inherited from the component $B_y$ on the magnetospheric side of the reconnecting current layer. This is much in the same way that IMF $B_y$, when it is strong, can directly supply a guide field to the flux ropes. We recall that the $\beta^\star = 0.5$ contour we use as a magnetopause proxy is in fact slightly on the magnetospheric side of the current sheet. So, the four-quadrant structure in $B_y$ at the $\beta^\star = 0.5$ is just what is expected for a dipole-like magnetospheric field.

\subsection{FAC generation}

Although the pulsed Earth-bound FACs generated near the magnetopause are observed when FTE O-lines pass by, our study does not determine whether the FAC-carrying field lines directly connect back to the FTE's coiled field structure. In studies of Hall reconnection, the field-aligned, Alfv\'enic currents are in fact generated near X-lines \cite{nagai_2003_structure_hall_current-system_reconnection, hoshino_2001_suprathermal_electrons_in_reconnection_PIC_simulations, dai_2017_kaws_and_hall_fields_in_reconnection}. Although some models of reconnection predict continual production of FACs, these currents may be transient as well \cite{artemyev_2018_facs_reconnection_satellite_observations}. 
So, it may be that the pulsed FACs are generated nearer to X-lines, which should pass a given point on the magnetopause at the same rate as the O-lines, in an alternate fashion. We looked into this issue briefly by marking the positions of the X-lines in the 2D cuts of Figs.~\ref{facs_propagation_fig}b,~\ref{facs_propagation_fig2}b---but the FAC pulses appeared to be more clearly associated with the O-lines than the X-lines (not shown). 
We also note that the dayside FACs may be generated by a process analogous to what occurs on the nightside; that is, the FACs may be generated as a result of the bulk flow patterns resulting from the reconnection \cite<e.g., >{workayehu_2025_vlasiator_BBFs, koikkalainen_2025_vlasiator_nightside_M-I_coupling_PREPRINT}.
A more detailed study of the topology of the magnetic field near the FTE is needed to understand the transient generation of the pulsed FACs and their relationship to the null lines.

It was shown in Fig.~4,~\cite{paul23}, that the magnetospheric cusp region exhibits heightened time-varying FACs when an FTE arrives, and an X-line can be observed at the FTE-cusp interface. This indicates the FACs may be generated by reconnection between the FTE and cusp field lines. However, the pulsed FAC events we observe occur on field lines slightly equatorward of the cusp---see curves $C_{345}$ and $C_{678}$ in Figs.~\ref{facs_propagation_fig}a,~\ref{facs_propagation_fig2}a. So, the pulsed FACs may derive from a different process.

\subsection{FTE Dynamics}

Although the formation of FTEs at the magnetopause is complex and multi-scale in nature, we can approximate the rate of FTE formation by counting the O-lines as they migrate across the magnetopause. From inspection of Figs.~\ref{facs_propagation_fig}b,~\ref{facs_propagation_fig2}b, we observe that approximately $\sim$10 FTEs cross the mid-latitude curves $C_{345}$ and $C_{678}$ over the course of $\sim1000\ \mathrm{s}$ in our simulation. So in this manner, we may estimate a coherent, long-lasting FTE passes near the cusp every $\sim100\ \mathrm{s}$ \cite<in agreement with >{pfau-kempf_2025_fte_evolution_vlasiator}. This is similar to the time scale of the large-amplitude geoelectric field fluctuations seen in Fig.~\ref{Exy_dByx_fig} and animation~S4. Other ground observations have suggested that FTEs may cause ionospheric fluctuations over $\sim$10~minute time scales \cite{glassmeierstellmacher96}---slightly longer but not radically different than the dynamical times in our simulation. The occurrence rate of FTEs depends naturally on their magnetospheric location as well as the solar wind conditions, e.g. as 2D Vlasiator simulations have shown that a significant IMF $B_x$ component can lead to differing FTE occurrence rates between the northern and southern hemispheres \cite{hoilijoki_2019_FTE_2d_vlasiator}. An analysis of the physical mechanism that governs the FTE formation and occurrence rate, e.g. the collisionless tearing mode \cite{daughton_2005_collisionless_tearing_theory_magnetopause}, is beyond the scope of this paper.

In the context of magnetospheric physics, the terms ``flux rope'' and ``flux transfer event'' are often used interchangeably. So, it is surprising to observe in our simulations that a single FTE can in fact be split into multiple flux ropes, at the 3D null point junctions that also mark reversals in the flux rope helicity. Magnetohydrodynamic simulations of Earth's magnetopause \cite<e.g., >{komar_2013_tracing_magnetic_separators, eggington22} and PIC simulations \cite{eggington22} have investigated 3D nulls at the Earth's magnetopause, but to our knowledge this splitting effect has not been reported previously. It is not immediately clear, for example, whether this phenomenon is unique to our simulation setup, whether it arises from a kinetic effect that cannot be observed in MHD, or whether it arises naturally from the ionosphere-magnetosphere coupling.  The 3D nulls, or alternatively the $J_\parallel$ O-line current junctions, appear near locations where the FTEs' field lines stop coiling around the central O-line, and instead are diverted towards the Earth. These so-called open-closed field line footpoints coincide at least in part with the Region~1 current system, and therefore may contribute significantly to FACs on the dayside. Because of this, it is plausible that the 3D nulls, by breaking up the FTE magnetic field topology, provide a route by which the FTE current can be supplied to the ionosphere.

\subsection{Geoelectric Field Structures}

The development of rotational E-field cells near $\mathrm{MLT}=12$, and their subsequent migration around the auroral oval (Fig.~\ref{E_polar_cap_fig}, animation S4), is reminiscent of the longitudinal motion of the FTE footpoints.
As shown for a few cases in \citeA{paul23}, and for a larger statistical sample in \citeA{pfau-kempf_2025_fte_evolution_vlasiator}, FTE footpoints tend to appear near noon, and move towards the nightside as the FTEs themselves flow past Earth. Because we do not trace the FTE motion other than to identify the instantaneous O-line locations, we cannot determine whether the FTE footpoints move around the auroral oval as in these prior studies, or rather poleward as suggested in  \citeA{omidi_2007_cusp_reconnection, daum_2008_FTE_m-i_coupling_observations_and_MHD_simulations}.
But, we may reasonably presume that the geographical boundaries and apparent azimuthal motion of the geoelectric E-field structures are related to the FTE footpoint downmapping and the motion of these footpoints along the auroral oval. In \citeA{paul23}, it was shown that multipolar FAC signatures can occur around FTE footpoints---since the geoelectric field is indirectly driven by the FACs, it is highly plausible that the geoelectric field signatures also follow the footpoint motion.

The rotational geoelectric field patterns we observe are suggestive of the rotational velocity and ionospheric current vector fields seen in TCVs. We interpret this to be a manifestation of Lenz's law. That is, the geoelectric field drives currents in Earth's crust, that should oppose the changes in the ionospheric currents. We note that downwards (upwards) FACs are associated with clockwise (counter-clockwise) ionospheric currents \cite[e.g.]{amm_2002_tcv_event_study}. So, for example, the pre-noon counter-clockwise rotational geoelectric field at time $t=1102~\mathrm{s}$ in Figure~\ref{E_polar_cap_fig} are an indication that the upwards FACs are weakening with time at this instant.

\subsection{Caveats and limitations}

Our simulations employ several simplifying assumptions: constant solar wind conditions and purely southward IMF, zero magnetic dipole tilt, a height-integrated ionospheric conductivity model, and a coupling scheme that maps FACs between the magnetosphere and ionosphere over large distances $\sim r_C$. The simulation represents the evolution of the magnetosphere-ionosphere system over a period of $\sim$27 minutes. Effects that depend strongly on the solar wind variability, dipole tilt, etc. are not captured by this study. So, our study is not intended to recreate any specific event, nor the average environmental conditions of this system, but rather to reveal its fundamental processes in a highly idealized yet realistic scenario. 

In order to obtain the geoelectric field from post-processing the simulation data, we applied two additional simplifying assumptions: 1) the horizontal ground magnetic field at Earth's surface can be computed by evaluating the Biot-Savart integral over the entire simulation volume, and 2) the surface conductivity $\sigma$ is constant throughout the Earth's surface.  The former assumption is appropriate in the (magnetic) ``quasistatic'' approximation \cite<e.g., >{jackson_book}, where currents vary slowly enough that the displacement current can be neglected. This is exactly the Darwin approximation used in our hybrid-kinetic solver \cite{palmroth25_vlasov_methods_living_review, ganse23}, which effectively neglects the retardation of the electromagnetic fields arising from the finite propagation speed of light \cite{krause_2007_darwin_approximation_vlasov}.
The latter assumption ignores how spatial inhomogeneity of $\sigma$ is known to influence the geoelectric field, e.g. for famous GIC events such as the Hydro-Qu\'ebec superstorm \cite{boteler19_1989_superstorm}. Recent work has shown that horizontal gradients in $\sigma$ can lead to sizable curl-free geoelectric fields \cite{juusola_2025_curl-free_E_at_conductivity_gradients}---such effects are not resolved in our study.

\section{Summary and Conclusions}\label{summary_sec}

This study highlights coordinated dynamics that occur in the magnetosphere and ionosphere, that ultimately induce a geoelectric field in the dayside of Earth's surface at high latitudes. This space-weather process has several actors: the FTEs, null lines, and 3D null points near the magnetopause; the propagating field-aligned currents in the magnetosphere; the closure of these currents in the ionosphere; and the interaction of this current system's time-varying magnetic field with Earth's conducting crust. The kinetic simulation studied here reveals the relationships between these various actors, and illuminates the causal chain of events.

We here summarize our main results, that contribute to the scientific understanding of this process:
\begin{enumerate}
\item Seemingly coherent FTEs may in fact be comprised of two or more distinct, magnetically-disconnected flux ropes. The topological separations between flux ropes are observed to occur at locations along an FTE O-line where the parallel current $J_\parallel$ is zero. The sense of the flux ropes' current helicity (LH or RH) naturally reverses with $\text{sign}(J_\parallel)$ across these junctions. We also observe that near the junctions, the wound-up FTE magnetic field, which generally wraps around the central O-line, can reroute towards Earth. In this arrangement, one end of the observed field lines is rooted in Earth's high-latitude ionosphere, and the other end extends to the solar wind. The junctions thus act as magnetopause ``holes'', as described in \cite{crooker_1990_mapping_FTEs_to_ionosphere}.
\item Along the null (X- and O-) lines, 3D magnetic nulls (where ${\bvec{B}=0}$) are found at the ${J_\parallel=0}$ junctions described above. In our highly symmetrical simulation, the 3D nulls are mostly found along the equator and noon meridian, but execute a constant process of creation and annihilation.
\item The sign of $J_\parallel$ at the null lines, and therefore the helicity of the FTEs' magnetic field, is highly correlated with $\text{sign}(B_y)$ of the underlying magnetopause. This results in a four-quadrant structure where $\text{sign}(J_\parallel)$ of the null lines changes across the equator as well as the noon meridian. We predict that this pattern may emerge during periods of steady solar wind when the IMF $B_y$ component is negligible.
\item Along certain magnetospheric field lines, pulsed field-aligned currents are observed to be generated near the magnetopause when an FTE passes by. Similarly to \citeA{wang_2019_facs_kaws_magnetopause, paul23}, we interpret these pulsed currents as being carried by shear Alfv\'en waves. These currents are apparent in the keograms of Figs.~\ref{facs_propagation_fig}b and \ref{facs_propagation_fig2}b, because $\text{sign}(J_\parallel)$ of the pulsed FACs is opposite from the background FACs on these field lines. The FACs propagate along the field line, from the magnetopause towards the coupling radius of our simulation, so that they ultimately map down to the ionosphere. In our simulations, these transient pulses of current map down near noon ($12 \pm 1$~MLT), just south of the Region~1 current system.
\item In the dayside auroral region, we observe a highly dynamic geoelectric field is induced, with magnitudes $\sim 0.1-0.2$~V/km. Pairs of rotational TCV-like geoelectric field structures are observed to form on either side of the noon meridian, after which they propagate in opposite directions along the auroral oval towards the nightside. We observe the development of these structures correlates with the arrival of the pulsed FACs in the dayside ionosphere. We interpret the handedness (clockwise or counterclockwise) to depend on the time-variation and handedness of the divergence-free ionospheric currents, via Lenz's law. The strongest dayside geoelectric fields in our simulation are found at latitudes $\lambda \sim80^\circ$ around MLT=12, which includes the footpoints of the FTE O-lines, and $\bvec{E}$ in this region is generally directed azimuthally.
\end{enumerate}

We conclude by remarking on the impacts of this work and on potential directions of future research. Our study relating FTEs to the geoelectric field has implications for the broader topic of space weather, because the geoelectric field can drive GICs in power lines and other grounded conductors. Although GICs are primarily of interest on Earth's nightside, where the magnetotail reconnection can lead to highly geoeffective substorms, significant currents can be induced on the dayside as well \cite{love_2023_1940_superstorm, pulkkinen_2004_halloween_storm_in_sweden}. Understanding the causal mechanisms that drive dayside GICs is therefore important for mitigating the impacts of space weather, in addition to its intrinsic scientific value. These multi-system processes could potentially be observed by upcoming satellite missions targeted at dayside magnetosphere-ionosphere coupling, such as The Tandem Reconnection and Cusp Electrodynamics Reconnaissance Satellites (TRACERS) and Solar wind Magnetosphere Ionosphere Link Explorer (SMILE). Hybrid codes such as \Vlasiator\ will be an effective tool for informing these missions \cite{lin_2025_review_global_hybrid_modeling_for_TRACERS}. Further observational and numerical work is needed to better understand the connection between FTEs and geoelectric fields demonstrated here: to determine how the coupling depends on varying solar wind and ground conditions, and to explore how these processes fit into the more general context of magnetosphere-ionosphere coupling.

\section*{Open Research}

\Vlasiator\ version 5.2 \cite{pfau-Kempf_2024_vlasiator_v5.2} is distributed under the GPL-2 open-source license. The output of \Vlasiator\ simulations are stored in a custom file format (see \url{github.com/fmihpc/vlsv}) which is readable with the open-source Analysator software \cite{battarbee21_analysator_v0.9}.  The data used for this study can be accessed via \citeA{suni_dataset_2024}.

\section*{Conflict of Interest}

The authors declare there are no conflicts of interest for this manuscript.

%
%
%
%



\acknowledgments

Vlasiator was developed with the European Research Council Starting grant 200141-QuESpace, and Consolidator grant 682068 – PRESTISSIMO. MP acknowledges the Research Council of Finland grants 336847, 374095, and 368539.
The Research Council of Finland supported the contributions of M.~Grandin (AERGELC'H project, grant 338629) and Y.~Pfau-Kempf (KIMCHI project, grant 339756). The work of JS was made possible by a doctoral researcher position at the Doctoral Programme in Particle Physics and Universe Sciences funded by the University of Helsinki.

MA acknowledges the Research Council of Finland grant numbers 352846 and 361901, and the Inno4Scale project via European High-Performance Computing Joint Undertaking (JU) under Grant Agreement No 101118139. The JU receives support from the European Union's Horizon Europe Programme. The work of MA is funded by the European Union (ERC grant WAVESTORMS - 101124500). Views and opinions expressed are however those of the author(s) only and do not necessarily reflect those of the European Union or the European Research Council Executive Agency. Neither the European Union nor the granting authority can be held responsible for them.

The authors thank the Finnish Computing Competence Infrastructure (FCCI), the Finnish Grid and Cloud Infrastructure (FGCI) and the University of Helsinki IT4SCI team for supporting this project with computational and data storage resources. The authors wish to acknowledge CSC – IT Center for Science, Finland, for computational resources. The simulation presented in this work was run on the LUMI-C supercomputer through the EuroHPC project Magnetosphere-Ionosphere Coupling in Kinetic 6D (MICK, project number EHPC-REG-2022R02-238).

The work of GC is supported by the Integration Fellowship of Le Studium Loire Valley Institute for Advanced Studies.

KH acknowledges Ari Viljanen for his expert input on the general topic of induced geoelectric fields.


%
%



\bibliography{paper_refs}

@ARTICLE{horaites23,
       author = {{Horaites}, {K.} and {Rintam{\"a}ki}, E. and {Zaitsev}, I. and {Turc}, L. and {Grandin}, M. and {Cozzani}, G. and {Zhou}, H. and {Alho}, M. and {Suni}, J. and {Kebede}, F. and {Gordeev}, E. and {George}, H. and {Battarbee}, M. and {Bussov}, M. and {Dubart}, M. and {Ganse}, U. and {Papadakis}, K. and {Pfau-Kempf}, Y. and {Tarvus}, V. and {Palmroth}, M.},
        title = "{Magnetospheric Response to a Pressure Pulse in a Three-Dimensional Hybrid-Vlasov Simulation}",
      journal = {\jgr (Space Physics)},
         year = 2023,
        month = aug,
       volume = {128},
       number = {8},
          eid = {},
        pages = {},
          doi = {10.1029/2023JA031374},
       adsurl = {https://ui.adsabs.harvard.edu/abs/2023JGRA..12831374H}
}

@article{ganse23,
    author = {Ganse, Urs and Koskela, Tuomas and Battarbee, Markus and Pfau-Kempf, Yann and Papadakis, Konstantinos and Alho, Markku and Bussov, Maarja and Cozzani, Giulia and Dubart, Maxime and George, Harriet and Gordeev, Evgeny and Grandin, Maxime and {Horaites}, {Konstantinos} and Suni, Jonas and Tarvus, Vertti and Kebede, Fasil Tesema and Turc, Lucile and Zhou, Hongyang and Palmroth, Minna},
    title = "{Enabling technology for global 3D + 3V hybrid-Vlasov simulations of near-Earth space}",
    journal = {\PoP},
    volume = {30},
    number = {4},
    year = {2023},
    month = {04},
    issn = {1070-664X},
    doi = {},
    url = {},
    note = {042902},
    eprint = {https://pubs.aip.org/aip/pop/article-pdf/doi/10.1063/5.0134387/16832177/042902\_1\_5.0134387.pdf},
}

@article{grandin23,
	author = {Grandin, Maxime and Luttikhuis, Thijs and Battarbee, Markus and Cozzani, Giulia and Zhou, Hongyang and Turc, Lucile and Pfau-Kempf, Yann and George, Harriet and Horaites, Konstantinos and Gordeev, Evgeny and Ganse, Urs and Papadakis, Konstantinos and Alho, Markku and Tesema, Fasil and Suni, Jonas and Dubart, Maxime and Tarvus, Vertti and Palmroth, Minna},
	title = {First 3D hybrid-Vlasov global simulation of auroral proton precipitation and comparison with satellite observations},
	DOI= "10.1051/swsc/2023017",
	url= "",
	journal = {J. Space Weather Space Clim.},
	year = 2023,
	volume = 13,
	pages = "20",
}

@ARTICLE{knight73,
       author = {{Knight}, Stephen},
        title = "{Parallel electric fields}",
      journal = {\planss},
         year = 1973,
        month = may,
       volume = {21},
       number = {5},
        pages = {741-750},
          doi = {10.1016/0032-0633(73)90093-7},
       adsurl = {https://ui.adsabs.harvard.edu/abs/1973P&SS...21..741K}
}

@article{shao02,
author = {Shao, X. and Guzdar, P. N. and Milikh, G. M. and Papadopoulos, K. and Goodrich, C. C. and Sharma, A. and Wiltberger, M. J. and Lyon, J. G.},
title = {Comparing ground magnetic field perturbations from global MHD simulations with magnetometer data for the 10 January 1997 magnetic storm event},
journal = {\jgr (Space Physics)},
volume = {107},
number = {A8},
pages = {SMP 11-1-SMP 11-10},
doi = {https://doi.org/10.1029/2000JA000445},
url = {},
eprint = {https://agupubs.onlinelibrary.wiley.com/doi/pdf/10.1029/2000JA000445},
year = {2002}
}

@article{welling20,
author = {Welling, Daniel T. and Love, Jeffrey J. and Rigler, E. Joshua and Oliveira, Denny M. and Komar, Colin M. and Morley, Steven K.},
title = {Numerical Simulations of the Geospace Response to the Arrival of an Idealized Perfect Interplanetary Coronal Mass Ejection},
journal = {Space Weather},
volume = {19},
number = {2},
pages = {},
doi = {https://doi.org/10.1029/2020SW002489},
url = {},
eprint = {https://agupubs.onlinelibrary.wiley.com/doi/pdf/10.1029/2020SW002489},
note = {},
year = {2021}
}

@article{boteler19_1989_superstorm,
author = {Boteler, D. H.},
title = {A 21st Century View of the March 1989 Magnetic Storm},
journal = {Space Weather},
volume = {17},
number = {10},
pages = {1427-1441},
doi = {https://doi.org/10.1029/2019SW002278},
url = {},
eprint = {https://agupubs.onlinelibrary.wiley.com/doi/pdf/10.1029/2019SW002278},
year = {2019}
}

@article{palmroth25_vlasov_methods_living_review,
author = {Palmroth, Minna and Ganse, Urs and Pfau-Kempf, Yann and Battarbee, Markus and Alho, Markku and Nättilä, Joonas and Zaitsev, Ivan and Cozzani, Giulia and Papadakis, Konstantinos and Kotipalo, Leo and Zhou, Hongyang and Turc, Lucile and Hoilijoki, Sanni and Grandin, Maxime and Pänkäläinen, Lauri and Sandroos, Arto and Alfthan, Sebastian},
year = {2025},
month = {11},
pages = {},
title = {Vlasov Methods in Space Physics and Astrophysics},
volume = {11},
journal = {Living Reviews in Computational Astrophysics},
doi = {10.1007/s41115-025-00024-0}
}

@misc{battarbee21_analysator_v0.9,
  author       = {Markus Battarbee and
                  Otto Akseli Hannuksela and
                  Yann Pfau-Kempf and
                  Sebastian von Alfthan and
                  Urs Ganse and
                  Riku Jarvinen and
                  Leo Kotipalo and
                  Jonas Suni and
                  Markku Alho and
                  Lucile Turc and
                  Ilja Honkonen and
                  Thiago Brito and
                  Maxime Grandin},
  title        = {fmihpc/analysator: January 2021 Release},
  year         = 2021,
  publisher    = {Zenodo},
  version      = {0.9},
  doi          = {10.5281/zenodo.4462515},
  url          = {},
  type = {Software}
}

@software{pfau-Kempf_2024_vlasiator_v5.2,
author = {Pfau-Kempf, Yann and Ganse, Urs and Battarbee, Markus and Kotipalo, Leo and Koskela, Tuomas and von Alfthan, Sebastian and Honkonen, Ilja and Alho, Markku and Sandroos, Arto and Papadakis, Konstantinos and Zhou, Hongyang and Palmu, Miro and Grandin, Maxime and Suni, Jonas and Tarvus, Vertti and Pokhotelov, Dimitry and Hokkanen, Jaro and Rossi, Tuomas and Lankinen, Juhana and Haaja, Veeti and Mendez Pratt, Audrey and Lalagüe, Arnaud and Horaites, Konstantinos},
doi = {10.5281/zenodo.3640593},
month = sep,
title = {{Vlasiator}},
url = {https://github.com/fmihpc/vlasiator},
year = {2024}
}

@article{eggington22,
author = {Eggington, J. W. B. and Desai, R. T. and Mejnertsen, L. and Chittenden, J. P. and Eastwood, J. P.},
title = {Time-Varying Magnetopause Reconnection During Sudden Commencement: Global MHD Simulations},
journal = {\jgr (Space Physics)},
volume = {127},
number = {4},
pages = {},
doi = {https://doi.org/10.1029/2021JA030006},
url = {},
eprint = {https://agupubs.onlinelibrary.wiley.com/doi/pdf/10.1029/2021JA030006},
note = {},
year = {2022}
}

@article{xu16,
author = {Xu, Shaosui and Liemohn, Michael W. and Dong, Chuanfei and Mitchell, David L. and Bougher, Stephen W. and Ma, Yingjuan},
title = {Pressure and ion composition boundaries at Mars},
journal = {\jgr (Space Physics)},
volume = {121},
number = {7},
pages = {6417-6429},
doi = {https://doi.org/10.1002/2016JA022644},
url = {},
eprint = {https://agupubs.onlinelibrary.wiley.com/doi/pdf/10.1002/2016JA022644},
year = {2016}
}

@ARTICLE{cagniard53,
       author = {{Cagniard}, Louis},
        title = "{Basic Theory of the Magneto-Telluric Method of Geophysical Prospecting}",
      journal = {Geophysics},
         year = 1953,
        month = jul,
       volume = {18},
       number = {3},
        pages = {605},
          doi = {10.1190/1.1437915},
       adsurl = {https://ui.adsabs.harvard.edu/abs/1953Geop...18..605C}
}

@article{pulkkinen06,
author = {Pulkkinen, Antti and Viljanen, Ari and Pirjola, Risto},
year = {2006},
month = {Aug.},
pages = {},
title = {Estimation of geomagnetically induced current levels from different input data},
volume = {4},
journal = {Space Weather--the International Journal of Research and Applications},
doi = {10.1029/2006SW000229}
}

@BOOK{jackson_book,
       author = {{Jackson}, J.~D.},
        title = "{Classical Electrodynamics, 3rd ed.}",
         year = 1999,
       adsurl = {https://ui.adsabs.harvard.edu/abs/1998clel.book.....J}
}

@article{love14,
author = {Love, Jeffrey J. and Swidinsky, Andrei},
title = {Time causal operational estimation of electric fields induced in the Earth's lithosphere during magnetic storms},
journal = {\grl},
volume = {41},
number = {7},
pages = {2266-2274},
doi = {https://doi.org/10.1002/2014GL059568},
url = {},
eprint = {https://agupubs.onlinelibrary.wiley.com/doi/pdf/10.1002/2014GL059568},
year = {2014}
}

@InCollection{visit,
 author = {Hank Childs and Eric Brugger and Brad Whitlock and Jeremy Meredith and Sean Ahern
    and David Pugmire and Kathleen Biagas and Mark Miller and Cyrus Harrison
    and Gunther H. Weber and Hari Krishnan and Thomas Fogal and Allen Sanderson
    and Christoph Garth and E. Wes Bethel and David Camp and Oliver R\"{u}bel
    and Marc Durant and Jean M. Favre and Paul Navr\'{a}til},
 doi = {10.1201/b12985},
 title = {VisIt: An End-User Tool For Visualizing and Analyzing Very Large Data},
 booktitle = {High Performance Visualization: Enabling Extreme-Scale Scientific Insight},
 pages = {357-372},
 month = {October},
 year = {2012}
}

@ARTICLE{alho24,
       author = {{Alho}, Markku and {Cozzani}, Giulia and {Zaitsev}, Ivan and {Tesema Kebede}, Fasil and {Ganse}, Urs and {Battarbee}, Markus and {Bussov}, Maarja and {Dubart}, Maxime and {Hoilijoki}, Sanni and {Kotipalo}, Leo and {Papadakis}, Konstantinos and {Pfau-Kempf}, Yann and {Suni}, Jonas and {Tarvus}, Vertti and {Workayehu}, Abiyot and {Zhou}, Hongyang and {Palmroth}, Minna},
        title = "{Finding Reconnection Lines And Flux Rope Axes Via Local Coordinates In Global Ion-Kinetic Magnetospheric Simulations}",
      journal = {\AnnGeo},
         year = 2024,
        month = may,
       volume = {42},
        pages = {145-161},
          doi = {10.5194/angeo-42-145-2024},
       adsurl = {https://ui.adsabs.harvard.edu/abs/2024AnGeo..42..145A}
}

@ARTICLE{paul23,
       author = {{Paul}, Arghyadeep and {Strugarek}, Antoine and {Vaidya}, Bhargav},
        title = "{Global-MHD Simulations Using MagPIE: Impact of Flux Transfer Events on the Ionosphere}",
      journal = {\jgr (Space Physics)},
         year = 2023,
        month = nov,
       volume = {128},
       number = {11},
          eid = {},
        pages = {},
          doi = {10.1029/2023JA031718},
archivePrefix = {arXiv},
       eprint = {2311.16134},
 primaryClass = {physics.space-ph},
       adsurl = {https://ui.adsabs.harvard.edu/abs/2023JGRA..12831718P}
}

@ARTICLE{fukushima76,
       author = {{Fukushima}, N.},
        title = "{Generalized theorem for no ground magnetic effect of vertical currents connected with Pedersen currents in the uniform-conductivity ionosphere}",
      journal = {Report of Ionosphere and Space Research in Japan},
         year = 1976,
        month = jun,
       volume = {30},
       number = {1-2},
        pages = {35-40},
       adsurl = {https://ui.adsabs.harvard.edu/abs/1976RISRJ..30...35F}
}

@ARTICLE{russellelphic78,
       author = {{Russell}, C.~T. and {Elphic}, R.~C.},
        title = "{Initial ISEE Magnetometer Results: Magnetopause Observations (Article published in the special issues: Advances in Magnetospheric Physics with GEOS- 1 and ISEE - 1 and 2.)}",
      journal = {\ssr},
         year = 1978,
        month = dec,
       volume = {22},
       number = {6},
        pages = {681-715},
          doi = {10.1007/BF00212619},
       adsurl = {https://ui.adsabs.harvard.edu/abs/1978SSRv...22..681R}
}

@article{southwood85,
title = {Theoretical aspects of ionosphere-magnetosphere-solar wind coupling},
journal = {Advances in Space Research},
volume = {5},
number = {4},
pages = {7-14},
year = {1985},
issn = {0273-1177},
doi = {https://doi.org/10.1016/0273-1177(85)90110-3},
url = {},
author = {D.J. Southwood}
}

@article{southwood87,
author = {Southwood, D. J.},
title = {The ionospheric signature of flux transfer events},
journal = {\jgr (Space Physics)},
volume = {92},
number = {A4},
pages = {3207-3213},
doi = {https://doi.org/10.1029/JA092iA04p03207},
url = {},
eprint = {https://agupubs.onlinelibrary.wiley.com/doi/pdf/10.1029/JA092iA04p03207},
year = {1987}
}

@article{haerendel78,
author = {Haerendel, G. and Paschmann, G. and Sckopke, N. and Rosenbauer, H. and Hedgecock, P. C.},
title = {The frontside boundary layer of the magnetosphere and the problem of reconnection},
journal = {\jgr (Space Physics)},
volume = {83},
number = {A7},
pages = {3195-3216},
doi = {https://doi.org/10.1029/JA083iA07p03195},
url = {},
eprint = {https://agupubs.onlinelibrary.wiley.com/doi/pdf/10.1029/JA083iA07p03195},
year = {1978}
}

@article{glassmeierstellmacher96,
title = {Mapping flux transfer events to the ionosphere},
journal = {Advances in Space Research},
volume = {18},
number = {8},
pages = {151-160},
year = {1996},
note = {The Three-dimensional Magnetosphere},
issn = {0273-1177},
doi = {https://doi.org/10.1016/0273-1177(95)00983-3},
url = {},
author = {K.-H Glassmeier and M Stellmacher}
}

@article{shi19,
  title = {Dimensionality, {{Coordinate System}} and {{Reference Frame}} for {{Analysis}} of {{In-Situ Space Plasma}} and {{Field Data}}},
  author = {Shi, Q. Q. and Tian, A. M. and Bai, S. C. and Hasegawa, H. and Degeling, A. W. and Pu, Z. Y. and Dunlop, M. and Guo, R. L. and Yao, S. T. and Zong, Q.-G. and Wei, Y. and Zhou, X.-Z. and Fu, S. Y. and Liu, Z. Q.},
  year = {2019},
  month = may,
  journal = {Space Science Reviews},
  volume = {215},
  number = {4},
  pages = {35},
  issn = {1572-9672},
  doi = {10.1007/s11214-019-0601-2},
  langid = {english}
}

@article{abda20,
  title={A review of geomagnetically induced current effects on electrical power system: Principles and theory},
  author={Abda, Zmnako Mohammed Khurshid and Ab Aziz, Nur Fadilah and Ab Kadir, Mohd Zainal Abidin and Rhazali, Zeti Akma},
  journal={IEEE Access},
  volume={8},
  pages={200237--200258},
  year={2020},
  publisher={IEEE}
}

@article{pirjola02,
Author = {Pirjola, R},
Title = {Review on the calculation of surface electric and magnetic fields and of
   geomagnetically induced currents in ground-based technological systems},
Journal = {Surveys in Geophysics},
Year = {2002},
Volume = {23},
Number = {1},
Pages = {71-90},
Month = {Jan.},
Publisher = {KLUWER ACADEMIC PUBL},
Address = {VAN GODEWIJCKSTRAAT 30, 3311 GZ DORDRECHT, NETHERLANDS},
Type = {},
Language = {English},
Affiliation = {Pirjola, R (Corresponding Author), Finnish Meteorol Inst, Geophys Res Div, POB 503, FIN-00101 Helsinki, Finland.
   Finnish Meteorol Inst, Geophys Res Div, FIN-00101 Helsinki, Finland.},
DOI = {10.1023/A:1014816009303},
ISSN = {0169-3298},
Keywords-Plus = {COMPLEX-IMAGE METHOD; POWER TRANSMISSION-SYSTEMS; PJM INTERCONNECTION
   SYSTEM; AURORAL ELECTROJET; EARTHS SURFACE; INDUCTION; SITUATIONS},
Research-Areas = {Geochemistry \& Geophysics},
Web-of-Science-Categories  = {Geochemistry \& Geophysics},
Affiliations = {Finnish Meteorological Institute},
Number-of-Cited-References = {62},
Times-Cited = {118},
Usage-Count-Last-180-days = {1},
Usage-Count-Since-2013 = {15},
Journal-ISO = {Surv. Geophys.},
Doc-Delivery-Number = {534JW},
Web-of-Science-Index = {Science Citation Index Expanded (SCI-EXPANDED)},
Unique-ID = {WOS:000174583200003},
DA = {2024-06-13},
}

@article{liu09,
author = {Liu, Chun-Ming and Liu, Lian-Guang and Pirjola, Risto and Wang, Ze-Zhong},
title = {Calculation of geomagnetically induced currents in mid- to low-latitude power grids based on the plane wave method: A preliminary case study},
journal = {Space Weather},
volume = {7},
number = {4},
pages = {},
doi = {https://doi.org/10.1029/2008SW000439},
url = {},
eprint = {https://agupubs.onlinelibrary.wiley.com/doi/pdf/10.1029/2008SW000439},
year = {2009}
}

@article{karimabadi99,
author = {Karimabadi, H. and Krauss-Varban, D. and Omidi, N. and Vu, H. X.},
title = {Magnetic structure of the reconnection layer and core field generation in plasmoids},
journal = {\jgr (Space Physics)},
volume = {104},
number = {A6},
pages = {12313-12326},
doi = {https://doi.org/10.1029/1999JA900089},
url = {},
eprint = {https://agupubs.onlinelibrary.wiley.com/doi/pdf/10.1029/1999JA900089},
year = {1999}
}

@article{dahani22,
author = {Dahani, S. and Kieokaew, R. and Génot, V. and Lavraud, B. and Chen, Y. and Michotte de Welle, B. and Aunai, N. and Tóth, G. and Cassak, P. A. and Fargette, N. and Fear, R. C. and Marchaudon, A. and Gershman, D. and Giles, B. and Torbert, R. and Burch, J.},
title = {The Helicity Sign of Flux Transfer Event Flux Ropes and Its Relationship to the Guide Field and Hall Physics in Magnetic Reconnection at the Magnetopause},
journal = {\jgr (Space Physics)},
volume = {127},
number = {11},
pages = {},
doi = {https://doi.org/10.1029/2022JA030686},
url = {},
eprint = {https://agupubs.onlinelibrary.wiley.com/doi/pdf/10.1029/2022JA030686},
note = {},
year = {2022}
}

@article{kieokaew21,
author = {Kieokaew, R. and Lavraud, B. and Fargette, N. and Marchaudon, A. and Génot, V. and Jacquey, C. and Gershman, D. and Giles, B. and Torbert, R. and Burch, J.},
title = {Statistical Relationship Between Interplanetary Magnetic Field Conditions and the Helicity Sign of Flux Transfer Event Flux Ropes},
journal = {Geophysical Research Letters},
volume = {48},
number = {6},
pages = {},
doi = {https://doi.org/10.1029/2020GL091257},
url = {},
eprint = {https://agupubs.onlinelibrary.wiley.com/doi/pdf/10.1029/2020GL091257},
note = {},
year = {2021}
}

@article{leefu85,
author = {Lee, L. C. and Fu, Z. F.},
title = {A theory of magnetic flux transfer at the Earth's magnetopause},
journal = {Geophysical Research Letters},
volume = {12},
number = {2},
pages = {105-108},
doi = {https://doi.org/10.1029/GL012i002p00105},
url = {},
eprint = {https://agupubs.onlinelibrary.wiley.com/doi/pdf/10.1029/GL012i002p00105},
year = {1985}
}

@article{wendel_adrian_2013_magnetic_nulls_cluster,
author = {Wendel, D. E. and Adrian, M. L.},
title = {Current structure and nonideal behavior at magnetic null points in the turbulent magnetosheath},
journal = {\jgr (Space Physics)},
volume = {118},
number = {4},
pages = {1571-1588},
doi = {https://doi.org/10.1002/jgra.50234},
url = {},
eprint = {https://agupubs.onlinelibrary.wiley.com/doi/pdf/10.1002/jgra.50234},
year = {2013}
}

@ARTICLE{lau_finn_1990_3d_magnetic_nulls_reconnection,
       author = {{Lau}, Yun-Tung and {Finn}, John M.},
        title = "{Three-dimensional Kinematic Reconnection in the Presence of Field Nulls and Closed Field Lines}",
      journal = {\apj},
         year = 1990,
        month = feb,
       volume = {350},
        pages = {672},
          doi = {10.1086/168419},
       adsurl = {https://ui.adsabs.harvard.edu/abs/1990ApJ...350..672L}
}

@article{greene_1988_properties_of_3d_magnetic_nulls_reconnection,
author = {Greene, John M.},
title = {Geometrical properties of three-dimensional reconnecting magnetic fields with nulls},
journal = {\jgr (Space Physics)},
volume = {93},
number = {A8},
pages = {8583-8590},
doi = {https://doi.org/10.1029/JA093iA08p08583},
url = {},
eprint = {https://agupubs.onlinelibrary.wiley.com/doi/pdf/10.1029/JA093iA08p08583},
year = {1988}
}

@article{murphy_2015_appearance_and_disappearance_of_3d_magnetic_nulls,
    author = {Murphy, Nicholas A. and Parnell, Clare E. and Haynes, Andrew L.},
    title = {The appearance, motion, and disappearance of three-dimensional magnetic null points},
    journal = {Physics of Plasmas},
    volume = {22},
    number = {10},
    pages = {102117},
    year = {2015},
    month = {10},
    issn = {1070-664X},
    doi = {10.1063/1.4934929},
    url = {},
    eprint = {https://pubs.aip.org/aip/pop/article-pdf/doi/10.1063/1.4934929/13529459/102117_1_online.pdf},
}

@article{van_driel-gesztelyi_2003_solar_helicity_observations,
title = {Observations of magnetic helicity},
journal = {Advances in Space Research},
volume = {32},
number = {10},
pages = {1855-1866},
year = {2003},
issn = {0273-1177},
doi = {https://doi.org/10.1016/S0273-1177(03)90619-3},
url = {},
author = {L. {van Driel-Gesztelyi} and P. Démoulin and C.H. Mandrini}
}

@article{russell_2019_current_magnetic_helicities_same_sign,
doi = {10.3847/1538-4357/ab40b4},
url = {},
year = {2019},
month = {oct},
publisher = {The American Astronomical Society},
volume = {884},
number = {1},
pages = {55},
author = {Russell, A. J. B. and Demoulin, P. and Hornig, G. and Pontin, D. I. and Candelaresi, S.},
title = {Do Current and Magnetic Helicities Have the Same Sign?},
journal = {The Astrophysical Journal}
}

@Article{pfau-kempf_2025_fte_evolution_vlasiator,
AUTHOR = {Pfau-Kempf, Y. and Papadakis, K. and Alho, M. and Battarbee, M. and Cozzani, G. and P\"ank\"al\"ainen, L. and Ganse, U. and Kebede, F. and Suni, J. and Horaites, K. and Grandin, M. and Palmroth, M.},
TITLE = {Global evolution of flux transfer events along the magnetopause from the dayside to the far tail},
JOURNAL = {Annales Geophysicae},
VOLUME = {43},
YEAR = {2025},
NUMBER = {2},
PAGES = {469--488},
URL = {},
DOI = {10.5194/angeo-43-469-2025}
}

@Article{juusola_2025_curl-free_E_at_conductivity_gradients,
AUTHOR = {Juusola, L. and Vanham\"aki, H. and Marshalko, E. and Kruglyakov, M. and Viljanen, A.},
TITLE = {Estimation of the 3-D geoelectric field at the Earth's surface using spherical elementary current systems},
JOURNAL = {Annales Geophysicae},
VOLUME = {43},
YEAR = {2025},
NUMBER = {1},
PAGES = {271--301},
URL = {},
DOI = {10.5194/angeo-43-271-2025}
}

@article{krause_2007_darwin_approximation_vlasov,
    author = {Krause, Todd B. and Apte, A. and Morrison, P. J.},
    title = {A unified approach to the Darwin approximation},
    journal = {Physics of Plasmas},
    volume = {14},
    number = {10},
    pages = {102112},
    year = {2007},
    month = {10},
    issn = {1070-664X},
    doi = {10.1063/1.2799346},
    url = {},
    eprint = {https://pubs.aip.org/aip/pop/article-pdf/doi/10.1063/1.2799346/15763188/102112_1_online.pdf},
}

@article{komar_2013_tracing_magnetic_separators,
author = {Komar, C. M. and Cassak, P. A. and Dorelli, J. C. and Glocer, A. and Kuznetsova, M. M.},
title = {Tracing magnetic separators and their dependence on IMF clock angle in global magnetospheric simulations},
journal = {\jgr (Space Physics)},
volume = {118},
number = {8},
pages = {4998-5007},
doi = {https://doi.org/10.1002/jgra.50479},
url = {},
eprint = {https://agupubs.onlinelibrary.wiley.com/doi/pdf/10.1002/jgra.50479},
year = {2013}
}

@Article{ganse_2025_vlasiator_ionosphere,
AUTHOR = {Ganse, U. and Pfau-Kempf, Y. and Zhou, H. and Juusola, L. and Workayehu, A. and Kebede, F. and Papadakis, K. and Grandin, M. and Alho, M. and Battarbee, M. and Dubart, M. and Kotipalo, L. and Lalag\"ue, A. and Suni, J. and Horaites, K. and Palmroth, M.},
TITLE = {The Vlasiator 5.2 ionosphere -- coupling a magnetospheric hybrid-Vlasov simulation with a height-integrated ionosphere model},
JOURNAL = {Geoscientific Model Development},
VOLUME = {18},
YEAR = {2025},
NUMBER = {2},
PAGES = {511--527},
URL = {},
DOI = {10.5194/gmd-18-511-2025}
}

@Article{koikkalainen_2025_vlasiator_nightside_M-I_coupling_PREPRINT,
AUTHOR = {Koikkalainen, V. and Grandin, M. and Kilpua, E. and Workayehu, A. and Zaitsev, I. and Juusola, L. and Tao, S. and Alho, M. and P\"ank\"al\"ainen, L. and Cozzani, G. and Horaites, K. and Suni, J. and Pfau-Kempf, Y. and Ganse, U. and Palmroth, M.},
TITLE = {Mapping transition region flows to the ionosphere in a global hybrid-Vlasov simulation (Pre-print)},
JOURNAL = {EGUsphere},
VOLUME = {2025},
YEAR = {2025},
PAGES = {1--24},
URL = {},
DOI = {10.5194/egusphere-2025-2265}
}

@Article{workayehu_2025_vlasiator_BBFs,
AUTHOR = {Workayehu, A. and Palmroth, M. and Grandin, M. and Juusola, L. and Alho, M. and Zaitsev, I. and Koikkalainen, V. and Horaites, K. and Pfau-Kempf, Y. and Ganse, U. and Battarbee, M. and Suni, J.},
TITLE = {Ionospheric signatures of a Bursty Bulk Flow in the 6D Vlasiator simulation},
JOURNAL = {Annales Geophysicae},
VOLUME = {43},
YEAR = {2025},
NUMBER = {2},
PAGES = {723--737},
URL = {},
DOI = {10.5194/angeo-43-723-2025}
}

@article{love_2023_1940_superstorm,
author = {Love, Jeffrey J. and Rigler, E. Joshua and Hartinger, Michael D. and Lucas, Greg M. and Kelbert, Anna and Bedrosian, Paul A.},
title = {The March 1940 Superstorm: Geoelectromagnetic Hazards and Impacts on American Communication and Power Systems},
journal = {Space Weather},
volume = {21},
number = {6},
pages = {},
doi = {https://doi.org/10.1029/2022SW003379},
url = {},
eprint = {https://agupubs.onlinelibrary.wiley.com/doi/pdf/10.1029/2022SW003379},
note = {},
year = {2023}
}

@article{pulkkinen_2004_halloween_storm_in_sweden,
author = {Pulkkinen, Antti and Lindahl, Sture and Viljanen, Ari and Pirjola, Risto},
title = {Geomagnetic storm of 29–31 October 2003: Geomagnetically induced currents and their relation to problems in the Swedish high-voltage power transmission system},
journal = {Space Weather},
volume = {3},
number = {8},
pages = {},
doi = {https://doi.org/10.1029/2004SW000123},
url = {},
eprint = {https://agupubs.onlinelibrary.wiley.com/doi/pdf/10.1029/2004SW000123},
year = {2005}
}

@article{lin_2025_review_global_hybrid_modeling_for_TRACERS,
  title={Global Hybrid Modeling in TRACERS Mission},
  author={Lin, Yu and Wang, Xueyi and Fuselier, Stephen A and Connor, Hyunju K and Miles, David M and Kletzing, Craig A},
  journal={Space Science Reviews},
  volume={221},
  number={6},
  pages={1--36},
  year={2025},
  publisher={Springer}
}

@article{nagai_2003_structure_hall_current-system_reconnection,
author = {Nagai, T. and Shinohara, I. and Fujimoto, M. and Machida, S. and Nakamura, R. and Saito, Y. and Mukai, T.},
title = {Structure of the Hall current system in the vicinity of the magnetic reconnection site},
journal = {\jgr (Space Physics)},
volume = {108},
number = {A10},
pages = {},
doi = {https://doi.org/10.1029/2003JA009900},
url = {},
eprint = {https://agupubs.onlinelibrary.wiley.com/doi/pdf/10.1029/2003JA009900},
year = {2003}
}

@article{hoshino_2001_suprathermal_electrons_in_reconnection_PIC_simulations,
author = {Hoshino, M. and Mukai, T. and Terasawa, T. and Shinohara, I.},
title = {Suprathermal electron acceleration in magnetic reconnection},
journal = {\jgr (Space Physics)},
volume = {106},
number = {A11},
pages = {25979-25997},
doi = {https://doi.org/10.1029/2001JA900052},
url = {},
eprint = {https://agupubs.onlinelibrary.wiley.com/doi/pdf/10.1029/2001JA900052},
year = {2001}
}

@article{dai_2017_kaws_and_hall_fields_in_reconnection,
author = {Dai, Lei and Wang, Chi and Zhang, Yongcun and Lavraud, Benoit and Burch, James and Pollock, Craig and Torbert, Roy B.},
title = {Kinetic Alfvén wave explanation of the Hall fields in magnetic reconnection},
journal = {Geophysical Research Letters},
volume = {44},
number = {2},
pages = {634-640},
doi = {https://doi.org/10.1002/2016GL071044},
url = {},
eprint = {https://agupubs.onlinelibrary.wiley.com/doi/pdf/10.1002/2016GL071044},
year = {2017}
}

@article{omidi_2007_cusp_reconnection,
author = {Omidi, N. and Sibeck, D. G.},
title = {Flux transfer events in the cusp},
journal = {Geophysical Research Letters},
volume = {34},
number = {4},
pages = {},
doi = {https://doi.org/10.1029/2006GL028698},
url = {},
eprint = {https://agupubs.onlinelibrary.wiley.com/doi/pdf/10.1029/2006GL028698},
year = {2007}
}

@article{artemyev_2018_facs_reconnection_satellite_observations,
author = {Artemyev, A. V. and Pritchett, P. L. and Angelopoulos, V. and Zhang, X.-J. and Nakamura, R. and Lu, S. and Runov, A. and Fuselier, S. A. and Wellenzohn, S. and Plaschke, F. and Russell, C. T. and Strangeway, R. J. and Lindqvist, P.-A. and Ergun, R. E.},
title = {Field-Aligned Currents Originating From the Magnetic Reconnection Region: Conjugate MMS-ARTEMIS Observations},
journal = {Geophysical Research Letters},
volume = {45},
number = {12},
pages = {5836-5844},
doi = {https://doi.org/10.1029/2018GL078206},
url = {},
eprint = {https://agupubs.onlinelibrary.wiley.com/doi/pdf/10.1029/2018GL078206},
year = {2018}
}

@article{marshalko_2021_geoelectric_field_3_approaches,
author = {Marshalko, Elena and Kruglyakov, Mikhail and Kuvshinov, Alexey and Juusola, Liisa and Kwagala, Norah Kaggwa and Sokolova, Elena and Pilipenko, Vyacheslav},
title = {Comparing Three Approaches to the Inducing Source Setting for the Ground Electromagnetic Field Modeling due to Space Weather Events},
journal = {Space Weather},
volume = {19},
number = {2},
pages = {},
doi = {https://doi.org/10.1029/2020SW002657},
url = {},
eprint = {https://agupubs.onlinelibrary.wiley.com/doi/pdf/10.1029/2020SW002657},
note = {},
year = {2021}
}

@article{elphic_1990_FTE_magnetopoause_ionosphere,
author = {Elphic, R. C. and Lockwood, M. and Cowley, S. W. H. and Sandholt, P. E.},
title = {Flux transfer events at the magnetopause and in the ionosphere},
journal = {Geophysical Research Letters},
volume = {17},
number = {12},
pages = {2241-2244},
doi = {https://doi.org/10.1029/GL017i012p02241},
url = {},
eprint = {https://agupubs.onlinelibrary.wiley.com/doi/pdf/10.1029/GL017i012p02241},
year = {1990}
}

@article{hoilijoki_2019_FTE_2d_vlasiator,
author = {Hoilijoki, S. and Ganse, U. and Sibeck, D. G. and Cassak, P. A. and Turc, L. and Battarbee, M. and Fear, R. C. and Blanco-Cano, X. and Dimmock, A. P. and Kilpua, E. K. J. and Jarvinen, R. and Juusola, L. and Pfau-Kempf, Y. and Palmroth, M.},
title = {Properties of Magnetic Reconnection and FTEs on the Dayside Magnetopause With and Without Positive IMF Bx Component During Southward IMF},
journal = {\jgr (Space Physics)},
volume = {124},
number = {6},
pages = {4037-4048},
doi = {https://doi.org/10.1029/2019JA026821},
url = {},
eprint = {https://agupubs.onlinelibrary.wiley.com/doi/pdf/10.1029/2019JA026821},
year = {2019}
}

@article{daum_2008_FTE_m-i_coupling_observations_and_MHD_simulations,
author = {Daum, P. and Wild, J. A. and Penz, T. and Woodfield, E. E. and Rème, H. and Fazakerley, A. N. and Daly, P. W. and Lester, M.},
title = {Global MHD simulation of flux transfer events at the high-latitude magnetopause observed by the Cluster spacecraft and the SuperDARN radar system},
journal = {\jgr (Space Physics)},
volume = {113},
number = {A7},
pages = {},
doi = {https://doi.org/10.1029/2007JA012749},
url = {},
eprint = {https://agupubs.onlinelibrary.wiley.com/doi/pdf/10.1029/2007JA012749},
year = {2008}
}

@article{wang_2019_facs_kaws_magnetopause,
    author = {Wang, Huanyu and Lin, Yu and Wang, Xueyi and Guo, Zhifang},
    title = {Generation of kinetic Alfvén waves in dayside magnetopause reconnection: A 3-D global-scale hybrid simulation},
    journal = {Physics of Plasmas},
    volume = {26},
    number = {7},
    pages = {072102},
    year = {2019},
    month = {07},
    issn = {1070-664X},
    doi = {10.1063/1.5092561},
    url = {},
    eprint = {https://pubs.aip.org/aip/pop/article-pdf/doi/10.1063/1.5092561/13819114/072102_1_online.pdf},
}

@misc{suni_dataset_2024,
    title = {[Dataset] Vlasiator {6D} '{FHA}' dataset},
    shorttitle = {Vlasiator {6D} '{FHA}' dataset},
    url = {},
    urldate = {2025-05-09},
    author = {Suni, J. and Horaites, K.},
    month = dec,
    year = {2024},
}

@article{crooker_1990_mapping_FTEs_to_ionosphere,
author = {Crooker, N. U. and G. L. Siscoe}, 
title = {On mapping flux transfer events to the ionosphere},
journal = {\jgr (Space Physics)},
volume = {95},
number = {A4},
pages = {3795-3799},
doi = {https://doi.org/10.1029/JA095iA04p03795},
url = {},
eprint = {https://agupubs.onlinelibrary.wiley.com/doi/pdf/10.1029/JA095iA04p03795},
year = {1990}
}

@article{vonalfthan_2014_new_vlasiator_simulations,
title = {Vlasiator: First global hybrid-Vlasov simulations of Earth's foreshock and magnetosheath},
journal = {Journal of Atmospheric and Solar-Terrestrial Physics},
volume = {120},
pages = {24-35},
year = {2014},
issn = {1364-6826},
doi = {https://doi.org/10.1016/j.jastp.2014.08.012},
url = {},
author = {S. {von Alfthan} and D. Pokhotelov and Y. Kempf and S. Hoilijoki and I. Honkonen and A. Sandroos and M. Palmroth}
}

@article{pfau-kempf_2020_dayside_3d_reconnection,
title = "Hybrid-Vlasov modeling of three-dimensional dayside magnetopause reconnection",
author = "Y. Pfau-Kempf and M. Palmroth and A. Johlander and L. Turc and M. Alho and M. Battarbee and M. Dubart and M. Grandin and U. Ganse",
year = "2020",
month = sep,
doi = "10.1063/5.0020685",
language = "English",
volume = "27",
journal = "Physics of Plasmas",
issn = "1070-664X",
publisher = "American Institute of Physics",
number = "9",
}

@article{fear_2017_flux_transferred_by_FTEs,
author = {Fear, R. C. and Trenchi, L. and Coxon, J. C. and Milan, S. E.},
title = {How Much Flux Does a Flux Transfer Event Transfer?},
journal = {\jgr (Space Physics)},
volume = {122},
number = {12},
pages = {12,310-12,327},
doi = {https://doi.org/10.1002/2017JA024730},
url = {},
eprint = {https://agupubs.onlinelibrary.wiley.com/doi/pdf/10.1002/2017JA024730},
year = {2017}
}

@ARTICLE{juusola_2020_internal_vs_external_B_field,
       author = {{Juusola}, Liisa and {Vanham{\"a}ki}, Heikki and {Viljanen}, Ari and {Smirnov}, Maxim},
        title = "{Induced currents due to 3D ground conductivity play a major role in the interpretation of geomagnetic variations}",
      journal = {Annales Geophysicae},
         year = 2020,
        month = sep,
       volume = {38},
       number = {5},
        pages = {983-998},
          doi = {10.5194/angeo-38-983-2020},
       adsurl = {https://ui.adsabs.harvard.edu/abs/2020AnGeo..38..983J}
}

@ARTICLE{schmucker_1970_induction_anomalies,
       author = {{Schmucker}, Ulrich},
        title = "{An Introduction to Induction Anomalies}",
      journal = {Journal of Geomagnetism and Geoelectricity},
         year = 1970,
        month = jan,
       volume = {22},
       number = {1-2},
        pages = {9-33},
          doi = {10.5636/jgg.22.9},
       adsurl = {https://ui.adsabs.harvard.edu/abs/1970JGG....22....9S}
}

@article{kuvshinov_2008_3D_global_geomagnetic_induction_model,
author = {Kuvshinov, Alexey},
year = {2008},
month = {03},
pages = {139-186},
title = {3-D Global Induction in the Oceans and Solid Earth: Recent Progress in Modeling Magnetic and Electric Fields from Sources of Magnetospheric, Ionospheric and Oceanic Origin},
volume = {29},
journal = {Surveys in Geophysics},
doi = {10.1007/s10712-008-9045-z}
}

@ARTICLE{pulkkinen_2003_SECS_Bint_Bext_separation,
       author = {{Pulkkinen}, Antti and {Amm}, Olaf and {Viljanen}, Ari and {BEAR working group}},
        title = "{Separation of the geomagnetic variation field on the ground into external and internal parts using the spherical elementary current system method}",
      journal = {Earth, Planets and Space},
         year = 2003,
        month = mar,
       volume = {55},
       number = {3},
        pages = {117-129},
          doi = {10.1186/BF03351739},
       adsurl = {https://ui.adsabs.harvard.edu/abs/2003EP&S...55..117P}
}

@ARTICLE{pirjola_1989_perfect_conductor_below_surface,
    author={Pirjola, R. and Viljanen, A.},
    journal={IEEE Transactions on Power Delivery}, 
    title={On geomagnetically-induced currents in the finnish 400 kV power system by an auroral electrojet current}, 
    year={1989},
    volume={4},
    number={2},
    pages={1239-1245},
    doi={10.1109/61.25609}}

@article{amm_2002_tcv_event_study,
author = {Amm, O. and Engebretson, M. J. and Hughes, T. and Newitt, L. and Viljanen, A. and Watermann, J.},
title = {A traveling convection vortex event study: Instantaneous ionospheric equivalent currents, estimation of field-aligned currents, and the role of induced currents},
journal = {Journal of Geophysical Research: Space Physics},
volume = {107},
number = {A11},
pages = {SIA 1-1-SIA 1-11},
doi = {https://doi.org/10.1029/2002JA009472},
url = {},
eprint = {https://agupubs.onlinelibrary.wiley.com/doi/pdf/10.1029/2002JA009472},
year = {2002}
}

@article{fu_2015_FOTE_magnetic_nulls_MMS,
author = {Fu, H. S. and Vaivads, A. and Khotyaintsev, Y. V. and Olshevsky, V. and André, M. and Cao, J. B. and Huang, S. Y. and Retinò, A. and Lapenta, G.},
title = {How to find magnetic nulls and reconstruct field topology with MMS data?},
journal = {Journal of Geophysical Research: Space Physics},
volume = {120},
number = {5},
pages = {3758-3782},
doi = {https://doi.org/10.1002/2015JA021082},
url = {},
eprint = {https://agupubs.onlinelibrary.wiley.com/doi/pdf/10.1002/2015JA021082},
year = {2015}
}

@article{fu_2018_magnetic_nulls_mms_observations,
author = {Fu, H. S. and Cao, J. B. and Cao, D. and Wang, Z. and Vaivads, A. and Khotyaintsev, Y. V. and Burch, J. L. and Huang, S. Y.},
title = {Evidence of Magnetic Nulls in Electron Diffusion Region},
journal = {Geophysical Research Letters},
volume = {46},
number = {1},
pages = {48-54},
doi = {https://doi.org/10.1029/2018GL080449},
url = {},
eprint = {https://agupubs.onlinelibrary.wiley.com/doi/pdf/10.1029/2018GL080449},
year = {2019}
}

@article{daughton_2005_collisionless_tearing_theory_magnetopause,
author = {Daughton, William and Karimabadi, H.},
title = {Kinetic theory of collisionless tearing at the magnetopause},
journal = {Journal of Geophysical Research: Space Physics},
volume = {110},
number = {A3},
pages = {},
doi = {https://doi.org/10.1029/2004JA010751},
url = {},
eprint = {https://agupubs.onlinelibrary.wiley.com/doi/pdf/10.1029/2004JA010751},
year = {2005}
}

@article{ngwira_2015_extreme_geolectric_fields,
author = {Ngwira, Chigomezyo M. and Pulkkinen, Antti A. and Bernabeu, Emanuel and Eichner, Jan and Viljanen, Ari and Crowley, Geoff},
title = {Characteristics of extreme geoelectric fields and their possible causes: Localized peak enhancements},
journal = {Geophysical Research Letters},
volume = {42},
number = {17},
pages = {6916-6921},
doi = {https://doi.org/10.1002/2015GL065061},
url = {},
eprint = {https://agupubs.onlinelibrary.wiley.com/doi/pdf/10.1002/2015GL065061},
year = {2015}
}

@ARTICLE{pulkkinen_2015_extreme_geoelectric_field_statistics,
       author = {{Pulkkinen}, Antti and {Bernabeu}, Emanuel and {Eichner}, Jan and {Viljanen}, Ari and {Ngwira}, Chigomezyo},
        title = "{Regional-scale high-latitude extreme geoelectric fields pertaining to geomagnetically induced currents}",
      journal = {Earth, Planets and Space},
         year = 2015,
        month = dec,
       volume = {67},
          eid = {93},
        pages = {93},
          doi = {10.1186/s40623-015-0255-6},
       adsurl = {https://ui.adsabs.harvard.edu/abs/2015EP&S...67...93P}
}

%
%
%
%
%

\end{document}